\documentclass[twocolumn,trackchanges]{aastex63}
\received{\today}
\revised{}
\accepted{}
\submitjournal{ApJ}

\shorttitle{}
\graphicspath{{./}{figures/}}

\begin{document}

\title{Multidimensional Boltzmann Neutrino Transport Code in Full General Relativity for Core-collapse Simulations \footnote{Released on June, 10th, 2019}}

\author[0000-0002-9234-813X]{Ryuichiro Akaho}
\affiliation{Graduate School of Advanced Science and Engineering,Waseda University,\\
3-4-1 Okubo, Shinjuku, Tokyo 169-8555, Japan}

\author[0000-0003-1409-0695]{Akira Harada}
\affiliation{Institute for Cosmic Ray Research, University of Tokyo,\\
5-1-5 Kashiwanoha, Kashiwa, Chiba 277-8582, Japan}

\author[0000-0002-7205-6367]{Hiroki Nagakura}
\affiliation{Department of Astrophysical Sciences, Princeton University,\\
Princeton, NJ 08544, USA}

\author[0000-0002-9224-9449]{Kohsuke Sumiyoshi}
\affiliation{National Institute of Technology, Numazu College,\\
Ooka 3600, Numazu, Shizuoka 410-8501, Japan}

\author[0000-0003-4959-069X]{Wakana Iwakami}
\affiliation{Advanced Research Institute for Science and Engineering, Waseda University, 3-4-1 Okubo, Shinjuku, Tokyo 169-8555, Japan}

\author{Hirotada Okawa}
\affiliation{Waseda Institute for Advanced Study, Waseda University, 1-6-1 Nishi-Waseda, Shinjuku-ku, Tokyo, 169-8050, Japan}

\author{Shun Furusawa}
\affiliation{Department of Physics, Tokyo University of Science, Shinjuku, Tokyo,
162-8601, Japan}

\author{Hideo Matsufuru}
\affiliation{High Energy Accelerator Research Organization, 1-1 Oho, Tsukuba, Ibaraki 305-0801, Japan}

\author[0000-0002-2166-5605]{Shoichi Yamada}
\affiliation{Advanced Research Institute for Science and Engineering, Waseda University,\\
3-4-1 Okubo, Shinjuku, Tokyo 169-8555, Japan}



\begin{abstract}
We develop a neutrino transfer code for core-collapse simulations, that directly solves the multidimensional Boltzmann equations in full general relativity. We employ the discrete ordinate method, which discretizes the six dimensional phase space. The code is an extension of our special relativistic code coupled to a Newtonian hydrodynamics code, which is currently employed for core-collapse supernova simulations. In order to demonstrate our code's capability to treat general relativistic effects, we conduct some tests: we first compute the free streaming of neutrinos in the Schwarzschild and Kerr spacetimes and compare the results with the geodesic curves; in the Schwarzschild case we deploy not only a 1-dimensional grid in space under spherical symmetry but also a 2-dimensional spatial mesh under axisymmetry in order to assess the capability of the code to compute the spatial advection of neutrinos; secondly, we calculate the neutrino transport in a fixed matter background, which is taken from a core-collapse supernova simulation with our general relativistic but spherically symmetric Boltzmann-hydrodynamics code, to obtain a steady neutrino distribution; the results are compared with those given by the latter code. 
\end{abstract}



\section{Introduction} 
\label{sec:intro}
Massive stars ($\gtrsim 8 M_{\odot}$) end their lives with gravitational collapse and eventually form compact objects such as neutron stars (NSs) and black holes (BHs) \citep[][for reviews]{Janka2012, Burrows2013, Foglizzo2015, OConnor2016}. Such a core-collapse event is a complicated phenomenon governed by nonlinear equations, and its accurate modeling is obtained only by numerical simulations. Ever since the first detailed study of the core-collapse events by \citet{Colgate1966}, many attempts have been made to model their dynamics and observational signals. Although successfully exploding models are not rare these days, the results do not agree with one another among different groups and, moreover, the explosion energy obtained by the numerical simulations is commonly small, only $\sim 10$\% of the typical observed values (\citet{OConnor2018a, Vartanyan2018, Burrows2019a, Burrows2019b, Muller2020, Kuroda2020}, but see also \citet{Burrows2020, Bollig2020}). Such discrepancies may be attributed to various approximations employed in the simulations, and there is a need for first-principles calculations, particularly of neutrino transfer without artificial approximations. 

The numerical handling of neutrino transport is indeed the important components for the core-collapse simulations. In fact, neutrinos carry most ($\sim 99\%$) of the energy released by the gravitational collapse, and the interactions between neutrinos and matter are believed to be crucial for the supernova dynamics \citep{Martinez-Pinedo2016,Halzen2017,Mezzacappa2020}: in modern simulations, the shock wave generated by the core bounce stalls on the way to the core surface because of the photo-dissociation of nuclei and most researchers think that the neutrino heating is the key process for shock revival to obtain successful explosions. 

Neutrinos are not in thermal equilibrium except deep inside the core and the neutrino distribution in momentum space needs to be calculated in principle. The neutrino transfer is described with the Boltzmann equations in the 6D phase space \citep{Lindquist1966, Ehlers1971}. There have been some attempts to directly solve them. The discrete ordinate ($S_N$) method \citep{Mezzacappa1993, Liebendorfer2001, Sumiyoshi2005, Sumiyoshi2012, Nagakura2014} discretizes the entire phase space and solve the equation with finite-difference. On the other hand, \citet{Chan2020} proposed ``intuitive particle-like approach'' for the calculation of the angular advection in momentum space, instead of the direct finite-differencing. Codes based on some spectral methods are also constructed: \citet{Radice2013} employs the spherical harmonics ($P_N$) to decompose the distribution function in momentum space whereas the distribution functions in both the configuration and momentum spaces are expanded with some basis functions by \citet{Peres2014}.

The Monte Carlo method \citep{Fleck1971, Abdikamalov2012, Richers2017} adopts a probabilistic description of neutrino transport. Monte Carlo transport is known to be noisy if the number of sample particles $N$ is not large enough, with the error scaling as $1/\sqrt{N}$. On the other hand, it can treat sharp angular distributions unlike the finite-difference schemes. \citet{Richers2017} made detailed comparisons between the Monte Carlo method and the discrete ordinate method and summarized the merits and demerits of the two methods.

Unfortunately, the Boltzmann transport schemes mentioned above are all very computationally demanding. Most core-collapse simulations have been hence performed by employing some phenomenological approximations: the flux-limited diffusion (FLD) approximation \citep{Arnett1977}, isotropic diffusion source approximation (IDSA) \citep{Liebendorfer2009}, ray-by-ray-plus (RbR+) approximation \citep{Buras2005}, and the truncated moment approximation \citep{Thorne1981, Shibata2011} with some analytical closure relations such as the M1 closure \citep{Levermore1984} being imposed by hand. Roughly speaking, these schemes reduce computational cost by neglecting some information on the distribution in momentum space. Several studies were conducted to compare these approximations \citep{Skinner2016, Cabezon2018, Just2018, OConnor2018b, Pan2018} and demonstrated that the different approximations may lead to quantitatively different results. This indicates that we need core-collapse simulations without appealing to these approximations, even if they are costly.

We should remind ourselves also that general relativity is another important ingredient. Core-collapse simulations with general relativistic neutrino transport, however, have been performed mostly with the approximate transport schemes \citep{Muller2010, Muller2012, Kuroda2012, Ott2013, Kuroda2016, Roberts2016, OConnor2018a, Kuroda2020}. In fact, Boltzmann simulations with the general relativistic discrete ordinate method are limited to spherical symmetry \citep{Yamada1999, Liebendorfer2001, Sumiyoshi2005}. Although Monte Carlo neutrino transport codes in full general relativity have been developed in the context of NS merger and collapsar simulations \citep{Foucart2016, Miller2019a, Miller2019b, Foucart2020}, they are yet to be applied to the core-collapse simulations.

The general relativistic radiative transport is a common problem for computational high-energy astrophysics. General relativistic photon transport schemes have been developed in the context of BH accretion and Monte Carlo photon transport codes have been constructed \citep{Dolence2009, Schnittman2013}. The long \citep[and references therein]{Takahashi2017} or short characteristics method \citep{Mihalas1978, Kunasz1988} is an alternative, which utilizes the transfer equation along the geodesic curves. Although the former is known to be very accurate, it is also computationally demanding and the latter is sometimes preferred. A hybrid method \citep{Zhu2015} is also developed.

We have developed a multidimensional Boltzmann-radiation hydrodynamics code, based on the discrete ordinate method. \citet{Sumiyoshi2012} developed a Newtonian transport code for the static background. It was later extended to incorporate special relativity to all orders of $v/c$ implementing the two-energy-grid scheme in \citet{Nagakura2014}. \citet{Nagakura2017} further extended the code to track the proper motion of a proto-neutron star by installing moving-coordinates technique that makes use of the gauge degree of freedom in general relativity, since the code employs the conservative form of the general relativistic Boltzmann equation written in the 3+1 decomposition formulation in \citet{Shibata2014}. Note, however, that this was realized only in the flat spacetime by a time--dependent uniform shift vector. We have run a series of CCSNe simulations with this code \citep{Nagakura2018, Nagakura2019, Harada2019, Harada2020, Iwakami2020} although they are limited to the flat spacetime.

In this paper, we report the results of some tests conducted for the fully general relativistic version of our code. We assess its capabilities of both the advection in phase space and the handling of interactions. Although the tests are limited to 1D (spherical symmetry) or 2D (axisymmetry) in this paper, our code is applicable to 3D calculations. Although we employ the spacetime metrics given analytically in this paper, it is applicable to any dynamical spacetimes such as those encountered in BH formations.

This paper is organized as follows. In section \ref{sec:formulation}, we describe the 3+1 formulation of the Boltzmann equation. The numerical method employed in our code is explained in section \ref{sec:numerical}. The results of the advection tests in the Schwarzschild and Kerr spacetimes are presented in sections \ref{sec:schwarzadv} and \ref{sec:kerrtest}, respectively. We then show the results of the tests on the collision terms in section \ref{sec:collision}. Finally, we summarize our findings and discuss future prospects in section \ref{sec:sum&dis}. Throughout the paper, we use the metric signature $-+++$; Greek ($\alpha,\beta,\mu,\nu$) and Latin ($i$, $j$, $k$) indices run over 0-3 and 1-3, respectively. 


\section{General Relativistic Formulation of Boltzmann Equation}
\label{sec:formulation}

The neutrino distribution in the curved spacetime is described by the general relativistic Boltzmann equation: \citep[for references]{Ehlers1971, Lindquist1966}
\begin{equation}
\label{eq:boltz}
p^\mu\frac{\partial f}{\partial x^\mu} - \Gamma^i_{\mu\nu}p^\mu p^\nu\frac{\partial f}{\partial p^i} = -p^\mu u_\mu S_\mathrm{rad},
\end{equation}
where $f$ is the distribution function in phase space, and $x^\alpha$ and $p^\mu$ are the spacetime coordinates and the 4-momentum, respectively. Note that the spatial components of the 4-momentum serve as the coordinates in momentum space; $\Gamma^\alpha_{\mu\nu}$ is the Christoffel symbol; $u^\mu$ is the four-vector of matter, and $S_\mathrm{rad}$ is the collision term; the first and second terms on the left-hand side of the equation describe the neutrino advection in space and momentum space, respectively. 

\begin{figure}[t]
  \includegraphics[width=9cm]{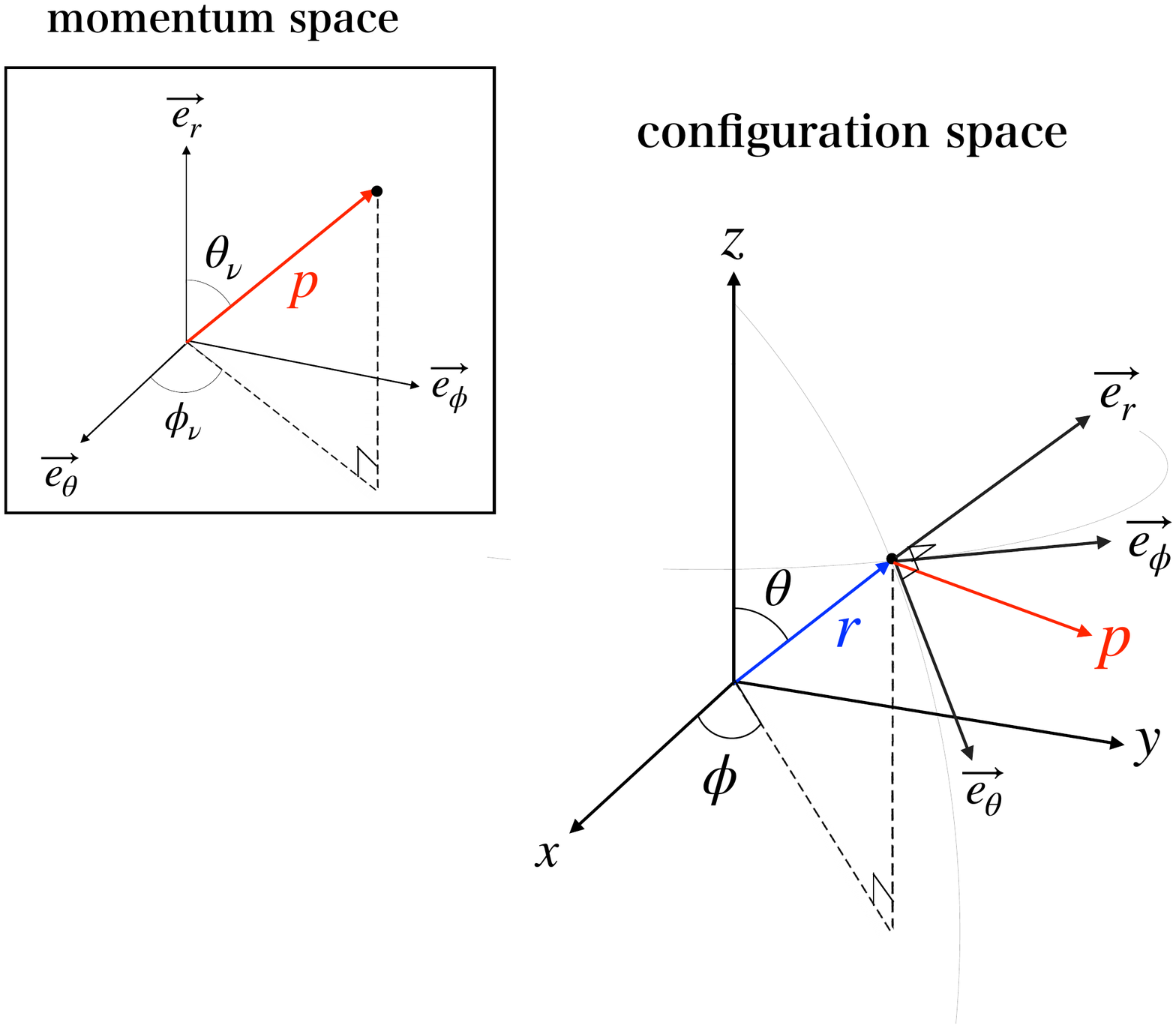}
  \caption{The schematic picture of the coordinate systems employed in this paper. For each point on the three-dimensional configuration space, there is a locally defined momentum space. The angles $\theta_\nu$ and $\phi_\nu$ represent the direction of neutrino momentum with respect to the local orthonormal bases $\vec{e_r}$, $\vec{e_\theta}$, and $\vec{e_\phi}$.}
  \label{Fig:phasespace}
\end{figure}

From a numerical point of view, it is desirable to cast the Boltzmann equation in the conservative form. \citet{Shibata2014} gave such a formulation to the general relativistic Boltzmann equation:
\begin{eqnarray}
\label{eq:conservBoltz}
&& \left.\frac{1}{\sqrt{-g}}\frac{\partial}{\partial x^\mu}\right|_{q_i}\left[\left(e_{(0)}^{\mu} + \sum_{i=1}^{3}l_i {e_{(i)}^\mu}\right)\sqrt{-g}f\right] \nonumber \\
&& -\frac{1}{\epsilon^2}\frac{\partial}{\partial\epsilon}\left(\epsilon^3f\omega_{(0)}\right) + \frac{1}{\mathrm{\mathrm{sin}\theta_\nu}}\frac{\partial}{\partial\theta_\nu}\left(\mathrm{sin}\theta_\nu f\omega_{(\theta_\nu)}\right) \nonumber \\
&& -\frac{1}{\mathrm{sin}^2\theta_\nu}\frac{\partial}{\partial{\phi_\nu}}\left(f\omega_{({\phi_\nu})}\right) = S_{\rm rad},
\end{eqnarray}
where the energy is defined as $\epsilon=-p_\mu e_{(0)}^\mu$; $\theta_\nu$ and $\phi_\nu$ are the zenith and azimuth angles in momentum space, respectively, as depicted in figure \ref{Fig:phasespace}; $g$ is the determinant of the metric tensor and the tetrad basis is specified by $e_{(\mu)}^\alpha$. In this paper, neutrinos are assumed to be massless with their minuscule masses being neglected. Directional cosines in momentum space $l_{(i)}$ are expressed as
\begin{eqnarray}
&& l_{(1)} = \mathrm{cos}\theta_\nu, \nonumber \\
&& l_{(2)} = \mathrm{sin}\theta_\nu\mathrm{cos}{\phi_\nu}, \nonumber \\
&& l_{(3)} = \mathrm{sin}\theta_\nu\mathrm{sin}{\phi_\nu}. \nonumber
\end{eqnarray}
The factors $\omega_{0}$, $\omega_{(\theta_\nu)}$, and $\omega_{(\phi_\nu)}$ are defined as
\begin{eqnarray}
&& \omega_{(0)} \equiv \epsilon^{-2} p^\mu p_\nu \nabla_\mu e_{(0)}^\mu, \nonumber \\
&& \omega_{(\theta_\nu)} \equiv \sum_{i=1}^{3}\omega_{(i)}\frac{\partial l_{(i)}}{\partial \theta_\nu}, \nonumber \\
&& \omega_{(\phi_\nu)} \equiv \sum_{i=2}^{3}\omega_{(i)}\frac{\partial l_{(i)}}{\partial \phi_\nu},  \nonumber \\
&& \omega_i \equiv \epsilon^{-2} p^\mu p_\nu \nabla_\mu e_{(i)}^\nu. \nonumber
\end{eqnarray}
A natural choice of the tetrad basis may be given as
\begin{eqnarray}
e^\mu_{(0)} && = n^\mu,
\nonumber \\
e^\mu_{(1)} && = 
\gamma_{rr}^{-1/2}\left(\frac{\partial}{\partial r}\right)^\mu,
\nonumber \\
e^\mu_{(2)} && = - \frac{\gamma_{r\theta}}{\sqrt{\gamma_{rr}(\gamma_{rr}\gamma_{\theta\theta} - \gamma_{r\theta}^2)}}
\left(\frac{\partial}{\partial r}\right)^\mu
\nonumber \\
&&
+ \sqrt{\frac{\gamma_{rr}}{\gamma_{rr}\gamma_{\theta\theta} - \gamma_{r\theta}^2}}
\left(\frac{\partial}{\partial\theta}\right)^\mu,
\nonumber \\
e^\mu_{(3)} && = 
\frac{\gamma^{r\phi}}{\sqrt{\gamma^{\phi\phi}}}
\left(\frac{\partial}{\partial r}\right)^\mu
+
\frac{\gamma^{\theta\phi}}{\sqrt{\gamma^{\phi\phi}}}
\left(\frac{\partial}{\partial \theta}\right)^\mu
\nonumber \\
&& +
\sqrt{\gamma^{\phi\phi}}
\left(\frac{\partial}{\partial \phi}\right)^\mu. \label{eq:tetrad}
\end{eqnarray}
Here $n^\mu$ and $\gamma_{ij}$ are the timelike unit vector normal to the time-constant hypersurface and the spatial metric on it, respectively. We choose this normal vector $n^{\mu}$ as the time-like base, since the 0th components of all the space-like bases are vanishing then; the two of the space-like bases are chosen so that they should span the meridian plane, with one of them being aligned with the radial coordinate; the last one is perpendicular to the meridian plane.
The components of the normal vector can be explicitly written as
\begin{equation}
n^\mu = (1/\alpha, -\beta^i/\alpha),
\end{equation}
where $\alpha$ is the lapse function and $\beta^i$ is the shift vector. In the 3+1 formulation, the line element is expressed as
\begin{eqnarray}
ds^2 = (-\alpha^2 + \beta^k\beta_k)dt^2
+ 2\beta_idtdx^i
+ \gamma_{ij}dx^idx^j.
\end{eqnarray}
In numerical computations, the values of the metric variables need to be evaluated both on the cell interfaces and centers. For the metrics given analytically as in sections \ref{sec:schwarzadv} and \ref{sec:kerrtest}, they are just evaluated from those analytic formulae either on the interfaces or at the centers. For the test in section \ref{sec:collision} in which the metric is given numerically only at the cell interfaces from another simulation, the cell center values are calculated by linear interpolation.

\section{Numerical Methods}
\label{sec:numerical}
\subsection{Extensions}
The code employed in this paper is an extended version of the code developed for the flat spacetime \citep{Sumiyoshi2012, Nagakura2014, Nagakura2017, Nagakura2019}. The major modifications are made to the advection terms. In appendix \ref{sec:advfindif}, we give the discretized version of the advection terms adopted in our general relativistic code. The collision terms, on the other hand, are essentially unchanged from the original code \citep{Sumiyoshi2012}. Since the advection terms are calculated in the laboratory frame and the collision terms in the fluid rest frame, Lorentz transformations between the two frames are required. The two-energy-grid technique \citep{Nagakura2014} is implemented in order to handle them. The general relativity does not affect it. Finally, the discretized equations are solved semi-implicitly: non-relativistic part of the advection terms are treated implicitly while relativistic part is treated explicitly, as described in \citet{Nagakura2014}. On the other hand, the collision terms are completely handled implicitly. For the matrix inversion, the Bi-CGSTAB method \citep{Saad2003} is utilized with the weighted point-Jacobi preconditioner \citep{Imakura2012}.

\subsection{Matter Background}
The original Boltzmann solver is already coupled to a Newtonian hydrodynamics solver and is currently used for CCSN simulations. In this paper, however, we fix the matter distribution in order to focus on the Boltzmann solver.

In sections \ref{sec:schwarzadv} and \ref{sec:kerrtest}, we compute the free streaming of neutrinos in vacuum. In actual simulations, we put a tenuous gas uniformly to avoid a complete vacuum, which could not be treated by the hydrodynamics code. Note that we need to feed something to the hydrodynamics module, which is tied with the Boltzmann code. The addition of the dilute gas does not affect the Boltzmann part at all, though. The fluid velocities are set to zero. This is important even if neutrino-matter interactions are all turned off, since the laboratory frame and the fluid rest frame coincide with each other then. Note that we store the neutrino distribution functions in the fluid-rest frame. In section \ref{sec:collision}, on the other hand, the background spacetime and hydrodynamic quantities are taken from a CCSN simulation as explained later.

\section{Advection Tests in the Schwarzschild Spacetime}
\label{sec:schwarzadv}
In this and next sections, we test our code's capability to treat neutrino advection both in space and momentum space. We switch off all neutrino reactions throughout these sections and hence the background gas plays no role. The $\beta$-parameter, which is defined in Appendix \ref{eq:beta} to control the amount
of upwinding in the numerical derivatives, becomes $\beta=1$ in this case. It means that the spatial advection terms are upwind-differenced completely (see appendix \ref{sec:advfindif} for detail). 

In this section we employ a Schwarzschild exterior solution, i.e., spherically symmetric stationary solution of the Einstein equation for vacuum, as the background spacetime. It is given as
\begin{eqnarray}
ds^2 =
&& -\left(1-\frac{2GM}{c^2r}\right)c^2dt^2 + \left(1-\frac{2GM}{c^2r}\right)^{-1}dr^2
\nonumber \\
&&
+ r^2(d\theta^2+\mathrm{sin}^2\theta d\phi^2),
\end{eqnarray}
where {$t$, $r$, $\theta$, and $\phi$} are the spacetime coordinates and $M$ is the mass of the central object; $c$ and $G$ are the light speed and the gravitational constant, respectively. The geodesic curves, along which non-interacting neutrinos move, in the Schwarzschild solution is described by a simple equation as given in appendix \ref{sec:schwarzgeo}. We numerically evaluate equation \ref{eq:tetrad} to obtain the tetrad components, which are analytically reduced in the present case to
\begin{equation}
e^0_{(0)} = \alpha^{-1}, \quad
e^1_{(1)} = \gamma_{rr}^{-1}, \quad
e^2_{(2)} = \frac{1}{r}, \quad
e^3_{(3)} = \frac{1}{r\mathrm{sin}\theta}. \label{eq:Schwarztetrad}
\end{equation}
We set the gravity source to be a sphere with the constant density of $\rho=1\times10^{15}\,{\rm g\,cm^{-3}}$ and a radius of $R=12\,{\rm km}$; Hence the mass is $M=3.62\,M_\odot$, whose Schwarzschild radius is $r=10.8\,{\rm km}$ and the photon sphere radius is $16\,{\rm km}$. Inside 12km we use the Schwarzschild interior solution for the uniform sphere of the same mass. The modification is not important, since we consider the neutrino propagation only outside this region in this section.

\subsection{Spherically Symmetric Tests}
\label{sec:1Dtest}
We present the results of advection tests under the assumption of spherical symmetry in the neutrino distribution function, i.e., it depends only on $r$, $\epsilon$, and $\theta_\nu$. Although our code is multi-dimensional, we suppress the angular degrees of freedom in space intentionally and treat the radial advection alone in space (plus the advections in momentum space) for the tests in section 4.1. In the following, we separately discuss the energy advection and the angular advection in momentum space in sections \ref{sec:eneadv} and \ref{sec:schwarzangadv}, respectively.

\subsubsection{Energy Advection Tests}
\label{sec:eneadv}
In the gravitational field, neutrinos change their energy as they move. In the computation of neutrino transport it is important to take such effects into account because the neutrino interactions strongly depend on the neutrino energy. 

In order to study the capability of our code to treat those energy changes, we conduct the following test. We fix the neutrino distribution function to $f=1$ on a single energy bin at a certain radius, which serves as a monochromatic and spherical neutrino source. We set $f=0$ elsewhere initially. Then neutrinos will flow out of this source and fill the space, the evolution of which we will compute with our new code. As for the initial angular distribution in momentum space, we assume that the neutrinos move in a single direction with the zenith angle $\theta_{\nu} = 0$ (outward) or $\pi$ (inward). We can test redshift in the former and blueshift in the latter. Note, however, that it is difficult in our code to set the single-angle distributions given above strictly. We hence set $f=1$ either on the first or on the last angular bin and $f=0$ on other bins in the numerical test. In order to focus on the energy advection, we switch off angular advection in this test.

The results are compared with the analytical formula for the neutrino energy $\epsilon_\mathrm{ana}$ as a function of radius:
\begin{equation}
\label{eq:eneana}
\epsilon_\mathrm{ana}(r) = \left(1-\frac{2GM}{c^2r_\mathrm{source}}\right) \left(1-\frac{2GM}{c^2r}\right)^{-1} \epsilon_\mathrm{source},
\end{equation}
where $\epsilon_\mathrm{source}$ and $r_\mathrm{source}$ are the energy and radius of the source, respectively. 

It should be stressed that for mesh-based codes like ours this is a very challenging problem, with sharp edges existing both in the energy and angular distributions. As will be witnessed later, rather large numerical diffusions occur inevitably. We choose this test, though, since this enables us to see most clearly if the code can reproduce the redshift/blueshift of neutrino energy as it moves in the gravitational well; the resolution dependence also manifests itself.

Throughout this test, we deploy the radial mesh with $N_r = 128$ grid points that covers the range $r\in[0,100]\,{\rm km}$. It is finer in the region $r\in [10,50]\,{\rm km}$. The number of angular mesh points in momentum space is $N_{\theta_\nu} = 20$. The energy mesh has logarithmically spaced grid points and covers the range $\epsilon\in[0,50]\,{\rm MeV}$. We vary the number of energy mesh points as $N_\epsilon = 20$, $30$, $40$, and $60$ to study the resolution dependence of the result. 

\begin{figure*}[t]
 \begin{center}
  \includegraphics[width=18cm]{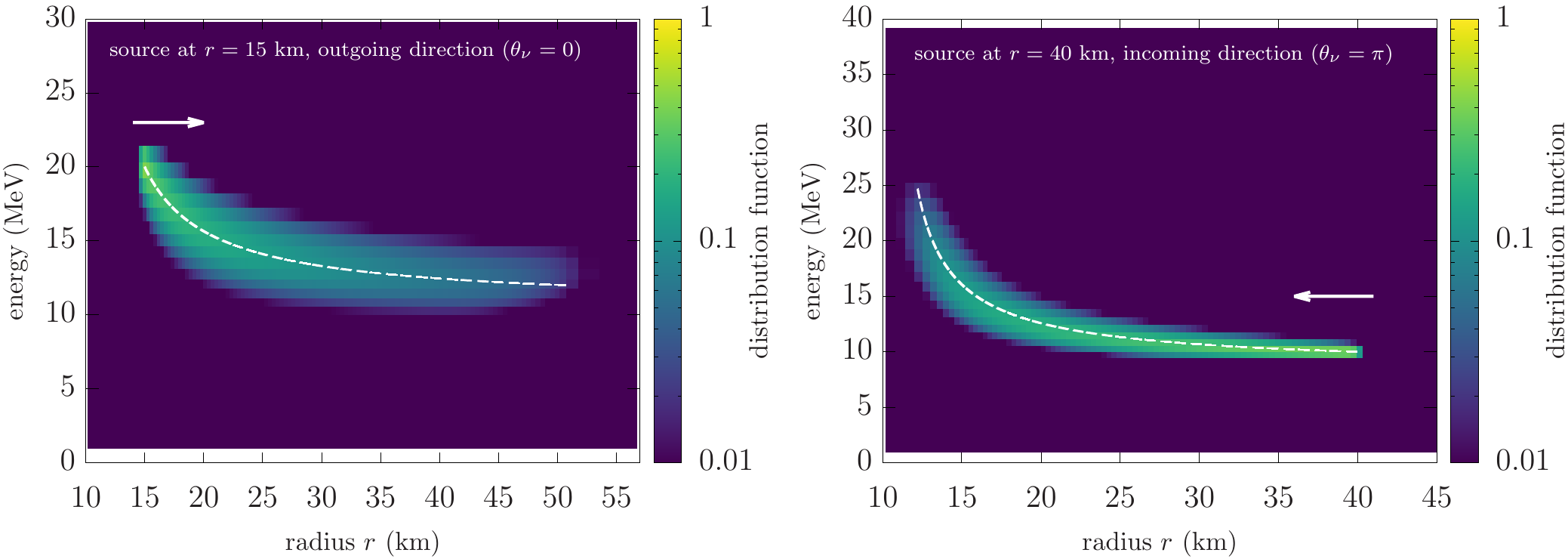}
  \caption{The neutrino distributions in energy space as a function of radius for the energy advection tests. The left and right panels show the results for the redshift and blueshift tests, respectively. The arrows indicate the directions of the neutrino motions, and the white dashed curves show the trajectory of the massless particles emitted from the source, truncated at the radius where the massless particles reach at the time of the snapshot $t=2\times10^{-4}\,{\rm s}$.}
  \label{Fig:eneshift}
 \end{center}
\end{figure*}

Figure \ref{Fig:eneshift} shows the energy distributions as a function of radius at a certain time for $N_\epsilon=60$. The white dashed curves depict the analytical solutions. The left panel presents the result of the gravitational redshift test at the coordinate time $t=2\times10^{-4}\,{\rm s}$. In this calculation, the neutrino source is located at $r=15\,{\rm km}$ and emits monochromatic neutrinos with $\epsilon=20\,{\rm MeV}$ outward. The energy distributions obtained numerically trace the analytical curve although they are somewhat broadened due to numerical diffusions arising from the finite energy resolution in the simulation. We stress again that this is actually a very challenging problem for finite difference methods like ours. In fact, the single energy bin has a finite width and cannot express the monochromatic energy distribution very well in the first place. The same trend is found for the blueshift test as exhibited in the right panel. The neutrino source is located at $r=40\,{\rm km}$ and emits neutrinos with $\epsilon=10\,{\rm MeV}$ inward in this test. The energy of neutrinos increases indeed as they propagate radially inward. It is also observed that the numerical diffusion is weaker at large radii as the advection is rather small there. In both panels, the analytic curves are drawn from the source positions to the points that the massless neutrino reaches at the given time. It is obvious that the terminal points are well reproduced by the numerical computations.

In order to see the resolution dependence of the numerical diffusion quantitatively, we repeat the redshift tests with different numbers of energy bins. We quantify the numerical diffusion by defining the following error function:
\begin{equation}
\label{eq:eneerr}
E_\epsilon(r)
\equiv
\frac{\sum_{n=1}^{N_\epsilon}f(r,\epsilon_n)(\epsilon_n - \epsilon_{\mathrm{ana}}(r))^2d\epsilon_n}
{({\epsilon_{\mathrm{ana}}(r))^2}\sum_{n=1}^{N_\epsilon}f(r,\epsilon_n)d\epsilon_n},
\end{equation}
where $\epsilon_n$ and $d\epsilon_n$ are the value of energy at the $n$-th cell center and the width of the same cell, respectively.

\begin{figure}[t]
  \includegraphics[width=8cm]{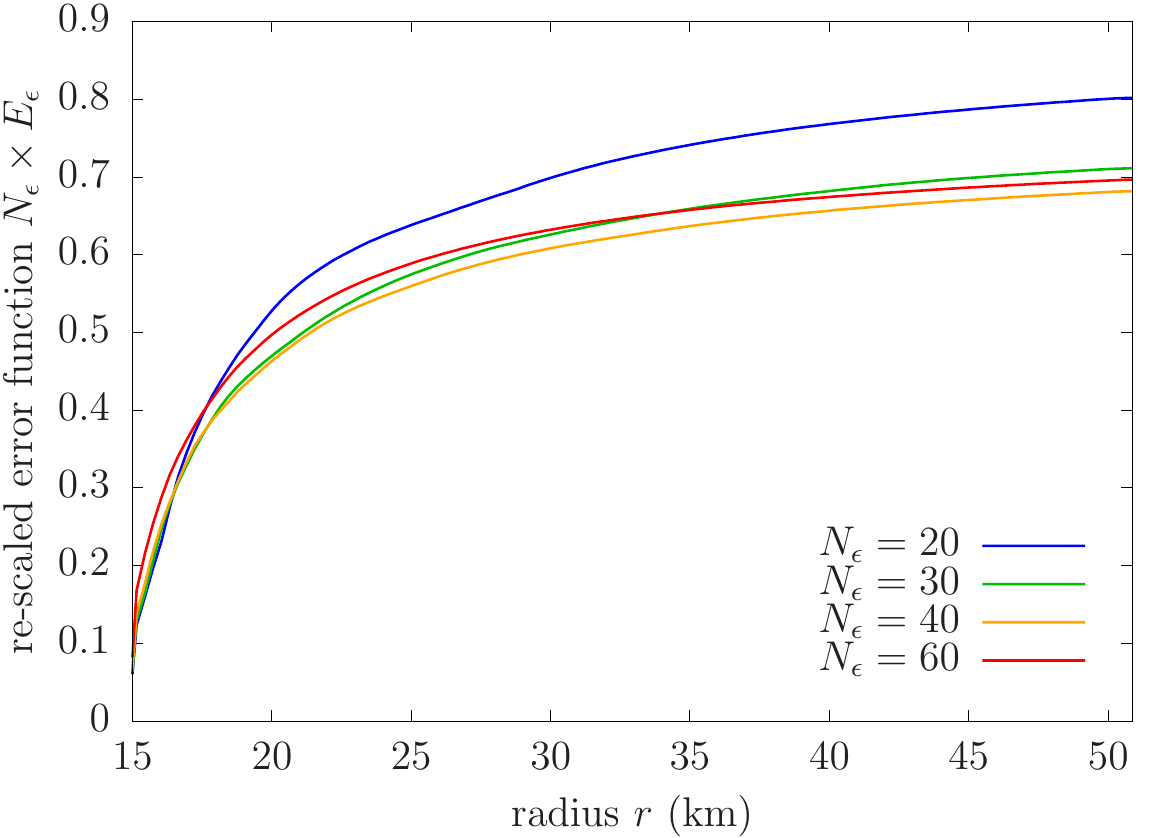}
  \caption{Radial profiles of the re-scaled error function defined in the text. Different colors indicate the numbers of energy mesh points: blue, green, yellow, and red curves are for $N_\epsilon=20$, $30$, $40$, and $60$, respectively.}
  \label{Fig:eneerr}
\end{figure}

In figure \ref{Fig:eneerr}, we show the radial profiles of the error function re-scaled by the number of energy mesh points $N_\epsilon$. It is seen that the re-scaled error functions for the different energy resolutions almost coincide with one another except for $N_{\epsilon}=20$. This indicates that the error function is inversely proportional to $N_\epsilon$, roughly implying the first-order convergence. This is expected, since the energy advection term is evaluated with a first-order finite difference scheme as described in appendix \ref{sec:advfindif}.

\subsubsection{Angular Advection Tests}
\label{sec:schwarzangadv}
\begin{figure}[t]
  \includegraphics[width=8cm]{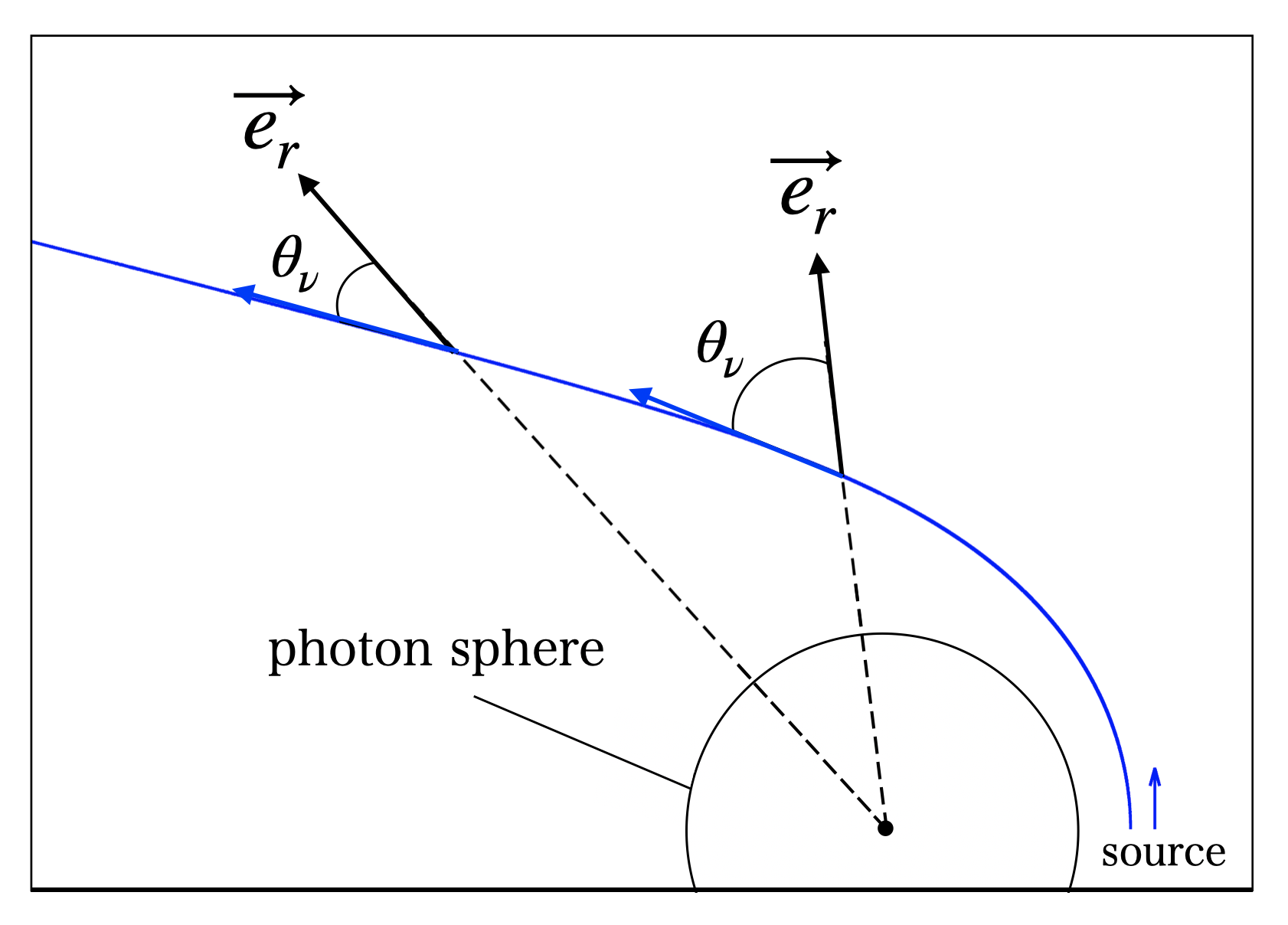}
  \caption{Schematic picture of the angular advection in momentum space angle $\theta_\nu$ for Schwarzschild spacetime. The blue curve indicates the trajectory of a massless particle emitted from the source located outside the photon sphere. The blue arrows are the tangent vectors of the trajectory; the black arrows are the radial vectors with the dashed lines indicating the radial ray from the coordinate center. The angle $\theta_\nu$ is the angle between these two vectors.}
  \label{Fig:schwarzang}
\end{figure}

The greatest advantage of directly solving the Boltzmann equation is that we are able to obtain information not only on energy but also on the angular distribution in momentum space. The direction of the neutrino momentum is specified by the zenith and azimuth angles, ($\theta_{\nu}$, $\phi_{\nu}$) (see figure \ref{Fig:phasespace}). Note that the distribution function depends on $\theta_{\nu}$ alone in the spherical symmetry assumed in this section. As a neutrino moves non-radially, the zenith angle $\theta_{\nu}$, which is measured from the local radial direction, changes even in the flat space time. This angular advection is shown schematically in Fig. \ref{Fig:schwarzang}. The blue curve is one of geodesic curves, along which the free neutrino moves in the Schwarzschild spacetime. Note that it is no longer a straight line due to gravity. In this example the neutrino moves outward and the zenith angle approaches $\theta_\nu=0$, i.e., the outward radial direction, with the increasing radius $r$. Since the geodesic curve is bent inward by gravity, the approach is slower for the Schwarzschild spacetime than in the flat spacetime. In this subsection, we test the capability of our code to reproduce this angular advection. 

The numerical setting is essentially the same as in the previous test for the energy advection: we put the monochromatic neutrino source uniformly on a sphere with a certain radius by setting $f=1$ on an single energy bin there and $f=0$ otherwise. The difference is that we choose $\theta_\nu=\pi/2$, which corresponds to direction of $p^r=0$, at the source. Note that we actually set $f=1$ on a single angular bin nearest to $\theta_{\nu} = \pi/2$ for numerical convenience. We vary the source radius to investigate the angular advections both inside and outside the photon sphere, i.e., the circular orbit. The neutrinos emitted outside the photon sphere with $p^r=0$ propagate outward with $\theta_\nu$ decreasing monotonically to zero whereas those emitted inside the photon sphere go inward with $\theta_{\nu}$ increasing as they propagate. In this test, we switch off the energy advection, which should occur simultaneously in reality. This is to avoid numerical diffusions both in angle and energy at the same time. The results are compared with those for the flat spacetime as well as with the reference solution obtained in appendix \ref{sec:schwarzgeo}.

We employ the same radial mesh with $N_r=128$ as in the previous tests. As for the mesh in momentum space, we vary the number of grid points as $N_{\theta_\nu}=20$, $30$ and $40$ to see the resolution dependence. We set $N_{\phi_{\nu}} = 2$ for numerical convenience although the distribution does not depend on $\phi_{\nu}$ in the present case.
 
\begin{figure*}[t]
 \begin{center}
  \includegraphics[width=18cm]{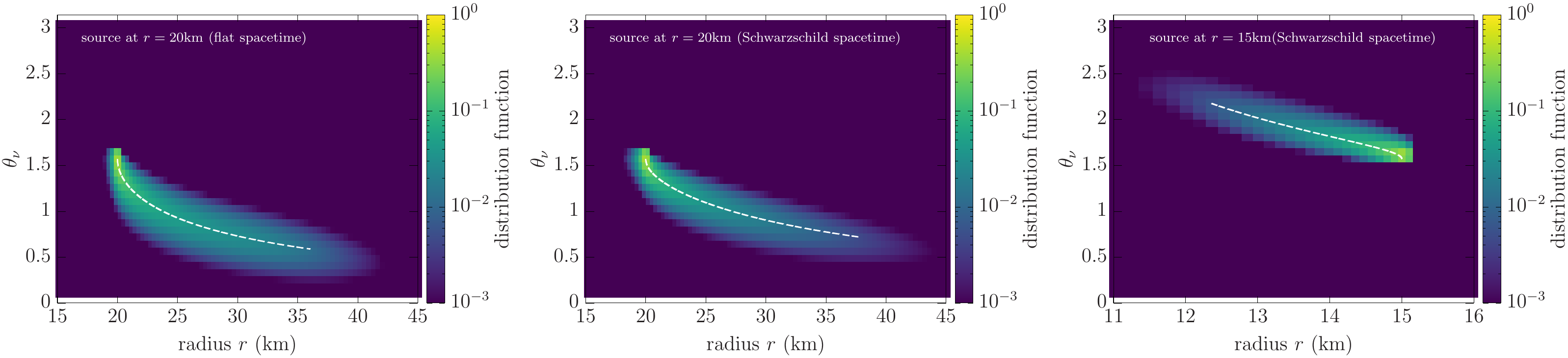}
  \caption{Angular distributions in momentum space as a function of the radius $r$. The white dashed curves show the reference geodesic curves, drawn from the source to the radius that the massless particles reach at the time of the snapshot. The left panel is the result for the flat spacetime at $t=1\times10^{-4}$ s, with the source placed at $r=20\,{\rm km}$. The middle and right panels are the results in the Schwarzschild spacetime at $t=2.5\times 10^{-4}\,{\rm s}$ and $t=2\times 10^{-4}\,{\rm s}$ with the sources located at $r=20\,{\rm km}$ (outside the photon sphere) and $r=15\,{\rm km}$ (inside the photon sphere), respectively.}
  \label{Fig:radang}
 \end{center}
\end{figure*}

\begin{figure}[t]
\begin{center}
  \includegraphics[width=8cm]{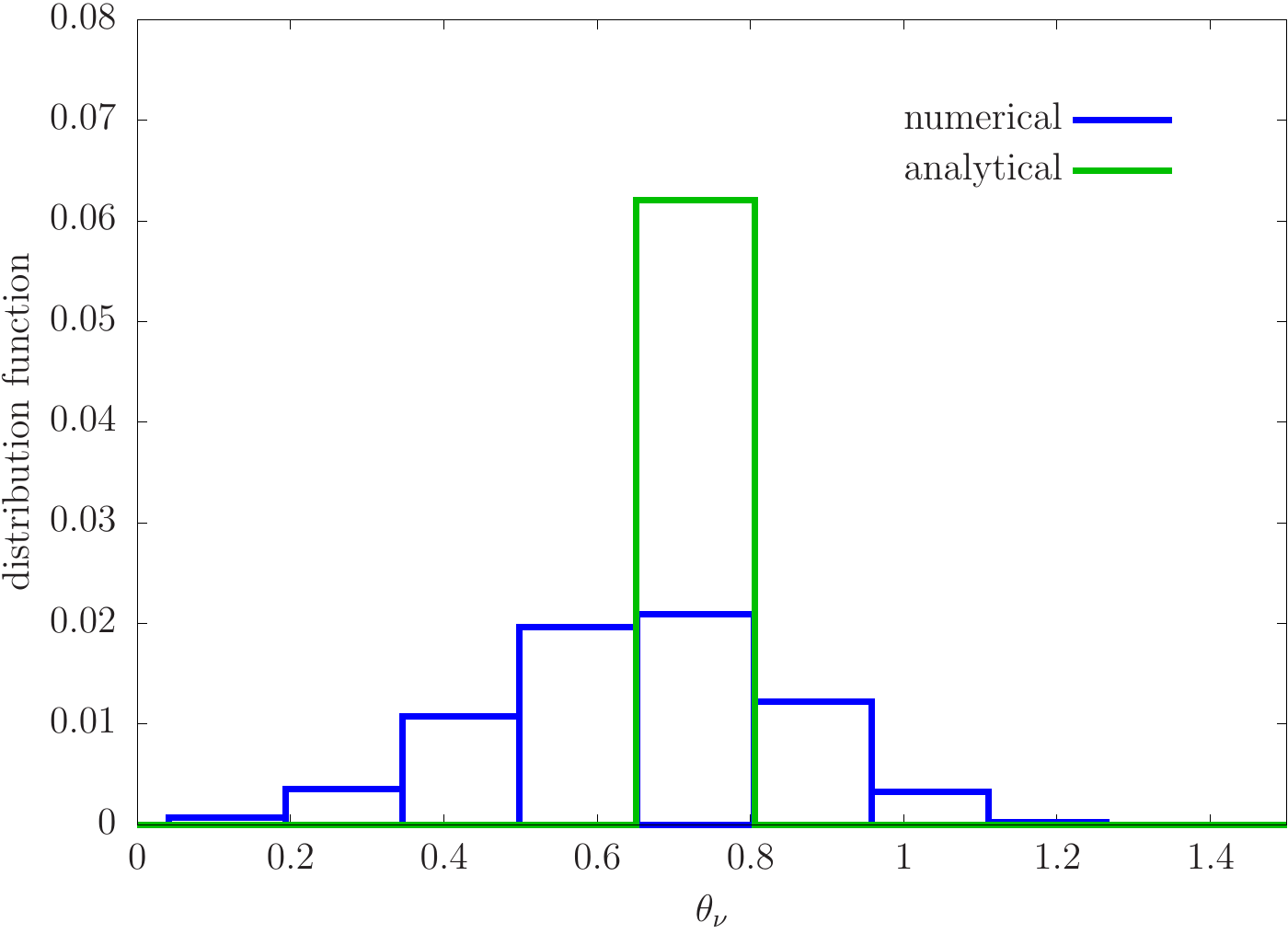}
  \caption{The angular distribution at $r=40\,{\rm km}$ for the angular advection test in the Schwarzschild spacetime. The green box indicates the analytical distribution, and the red box indicates the numerical result obtained with $N_{\theta_\nu}=20$. The snapshot is at the time $t=4\times10^{-4}\,{\rm s}$ when the distribution is stationary.}
  \label{Fig:tophat}
  \end{center}
\end{figure}

Figure \ref{Fig:radang} shows the angular distributions of neutrinos in momentum space as a function of the radius $r$ for the angular advection tests with $N_{\theta_\nu}=40$. The white dashed curves depict the reference geodesic curves, truncated at the radius that the massless particles reach at the time of the snapshot. The left panel shows the result for the flat spacetime at $t=1\times10^{-4}\,{\rm s}$, with the source placed at $r=20\,{\rm km}$. As mentioned earlier, there is a nontrivial angular advection even in the flat spacetime, since we deploy the polar coordinates in space. Neutrinos emitted with $p_r=0$ always move outward in the flat spacetime and the zenith angle monotonically converges to $\theta_\nu=0$ as they go outward. The middle panel presents the result at $t=2.5\times10^{-4}\,{\rm s}$ for the Schwarzschild spacetime. Note that the neutrino source is located outside the the photon sphere, the radius of which is $16\,{\rm km}$ in the present test. The neutrinos emitted from this source with $p_r = 0$ move outward. The outward radial propagation of neutrinos is slower in this case than in the flat case because of the geodesic deflection and the gravitational time delay. As a result, the radius-angle curve for the Schwarzschild spacetime is less steep than that for the flat spacetime. The right panel is the result at $t=2\times10^{-4}\,{\rm s}$ with the neutrino source located at $r=15\,{\rm km}$, i.e., inside the photon sphere in the Schwarzschild spacetime. This time, the trajectory is directed radially inward and, as a result, $\theta_{\nu}$ approaches $\theta_\nu = \pi$, instead of 0. The distributions are consistent with the reference curves although there are some numerical diffusions. Figure \ref{Fig:tophat} shows the radial slice of angular distribution at $r=40\,{\rm km}$ for the test in Schwarzschild spacetime. The numerical result obtained with $N_{\theta_\nu}=20$ is shown in red, where the analytical value is shown with green box. The height of the peak is lowered due to the numerical diffusion, however the position of the peak is consistent with the analytical value.

Just as in the energy advection tests in section \ref{sec:eneadv}, we quantify the numerical diffusion as follows: we define the error function as
\begin{equation}
\label{eq:angerr}
E_{\theta_\nu}(r)\equiv\frac{
 \sum_{n=1}^{N_{\theta_\nu}}f(r,(\theta_\nu)_n)((\theta_\nu)_n - (\theta_\nu)_{\mathrm{ref}})^2(d\theta_\nu)_n}
{\sum_{n=1}^{N_{\theta_\nu}}f(r,(\theta_\nu)_n)d\theta_\nu},
\end{equation}
where $(\theta_\nu)_n$ and $(d\theta_\nu)_n$ are the value of $\theta_\nu$ at the $n$-th cell center and the width of the same cell, respectively; $(\theta_\nu)_{\mathrm{ref}}$ is the zenith angle for the reference geodesic curve. We evaluate this function both for the flat and Schwarzschild spacetimes.

\begin{figure*}[t]
  \includegraphics[width=18cm]{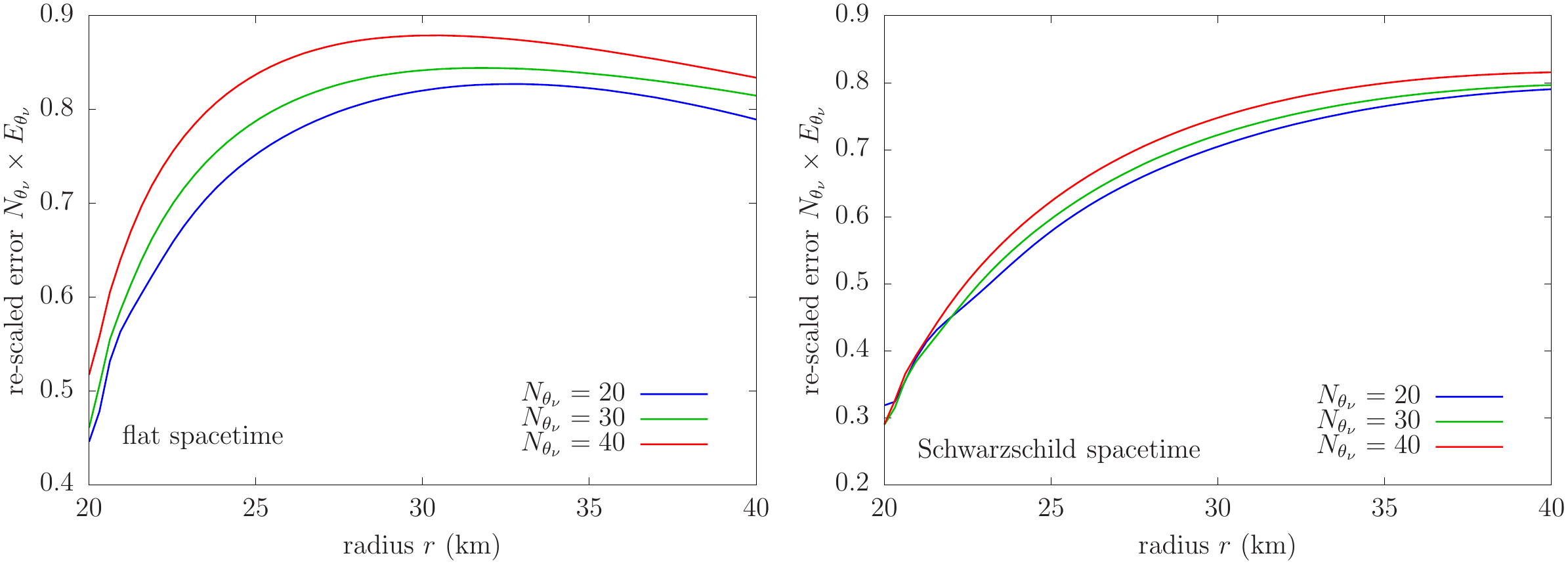}
  \caption{Radial profiles of the re-scaled error function defined in the text. The left and right panels are for the flat and Schwarzschild spacetimes, respectively. The blue, green, and red lines correspond to the different angular resolutions in momentum space: $N_{\theta_\nu} = 20$, $30$, and $40$.}
  \label{Fig:angerr} 
\end{figure*}

Figure \ref{Fig:angerr} shows the radial profiles of the re-scaled error function, i.e., the error function multiplied by the number of angular mesh points $N_{\theta_{\nu}}$. The left and right panels show the results for the flat and Schwarzschild spacetimes, respectively. For both cases, the re-scaled errors for the different numbers of mesh points are close to each other, implying that our code is of first order in the angular advection. This is just as expected, since we adopt the first-order finite difference scheme for the angular advection terms.

\subsubsection{Tests of Energy and Angular Advection Combined}
\begin{figure}[t]
  \includegraphics[width=8cm]{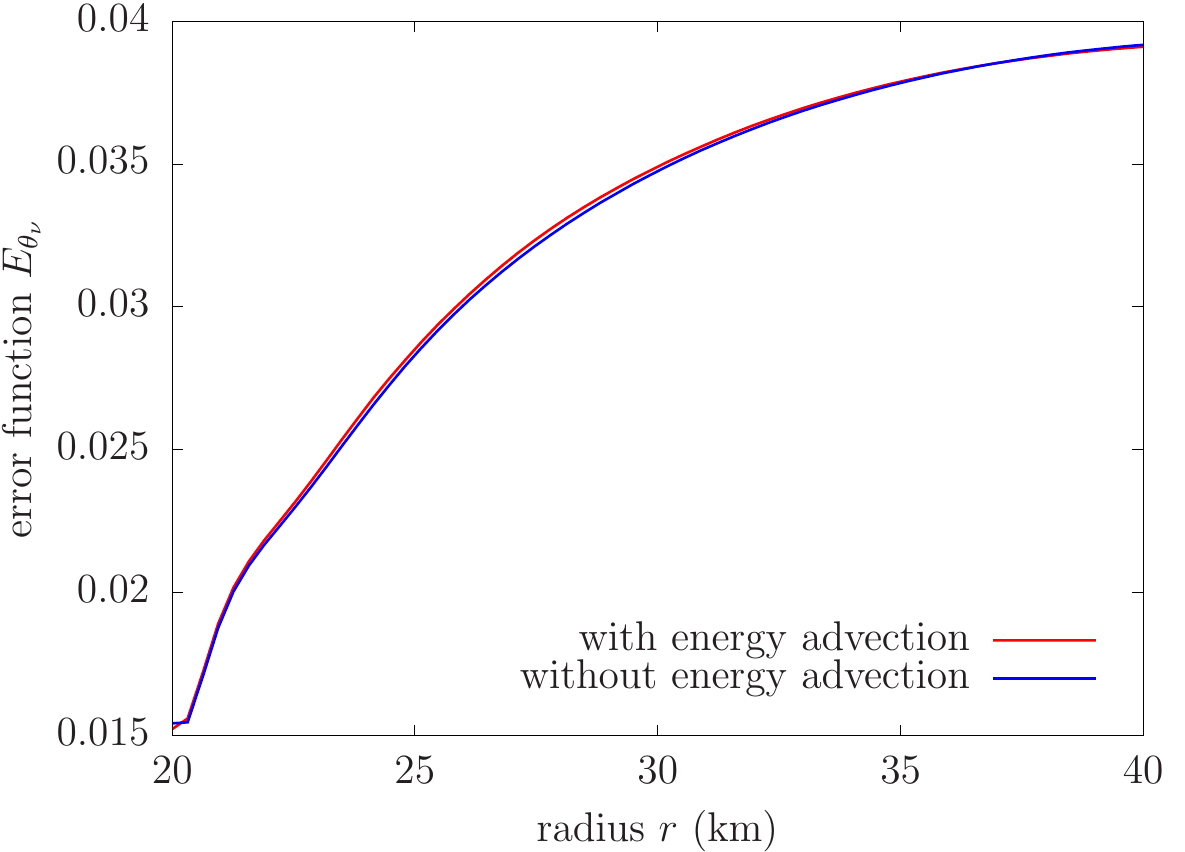}
  \caption{Radial profiles of the error function for angular advection. The red curve indicates the result with both angular and energy advection taken into account, and the blue curve is the one only with angular advection, same as \ref{sec:schwarzangadv}.}
  \label{Fig:angeneerr} 
\end{figure}

In this section, we conduct the advection tests, taking both the energy and angular advections into account, since they occur simultaneously in reality. Firstly, we re-do the same test as in section \ref{sec:schwarzangadv} with the energy advection turned on. We deploy the same mesh with $N_r=128$, $N_\epsilon=20$ and $N_{\theta_\nu}=20$ and place neutrino steady source at $r=20\,\mathrm{km}$ that emits neutrinos in a single angular bin closest to $\theta_\nu=\pi/2$ and a single energy bin at $\epsilon=20\,{\rm MeV}$. The angular advection error $E_{\theta_\nu}$ is calculated for the energy-integrated distribution function, and compared with the previous result without energy advection (see section \ref{sec:schwarzangadv}). Figure \ref{Fig:angeneerr} shows $E_{\theta_{\nu}}$ as a function of radius for both cases. The results are very close to each other, implying that the inclusion of the energy advection does not affect the angular advection.

Next, we consider a neutrino source that has smooth, extended energy and angular distributions in order to demonstrate that the numerical diffusion behavior is reduced for smoother distributions. We set the following distribution at the source:
\begin{equation}
f(x) = \left\{
\begin{array}{ll}
\frac{1}{1+e^{(\epsilon-\mu)/kT}}\frac{1+\mathrm{cos}2\theta_\nu}{2} & (\theta_\nu \leq \frac{\pi}{2}) \\
0 & (\theta_\nu > \frac{\pi}{2})
\end{array}
\right.,
\end{equation}
where we choose the parameters as $\mu=20\,{\rm MeV}$ and $kT=10\,{\rm MeV}$. We place this steady source at $r=20\,{\rm km}$. We employ the same radial mesh with $N_r=128$ as in the previous tests whereas we vary the number of grid points in the energy mesh that covers the range $\epsilon\in[0,300]\,{\rm MeV}$ as $N_{\epsilon} = 10$ and $20$; the cell number in the angular mesh is chosen to be either $N_{\theta_\nu}=20$ or $40$.

\begin{figure*}[t]
  \includegraphics[width=18cm]{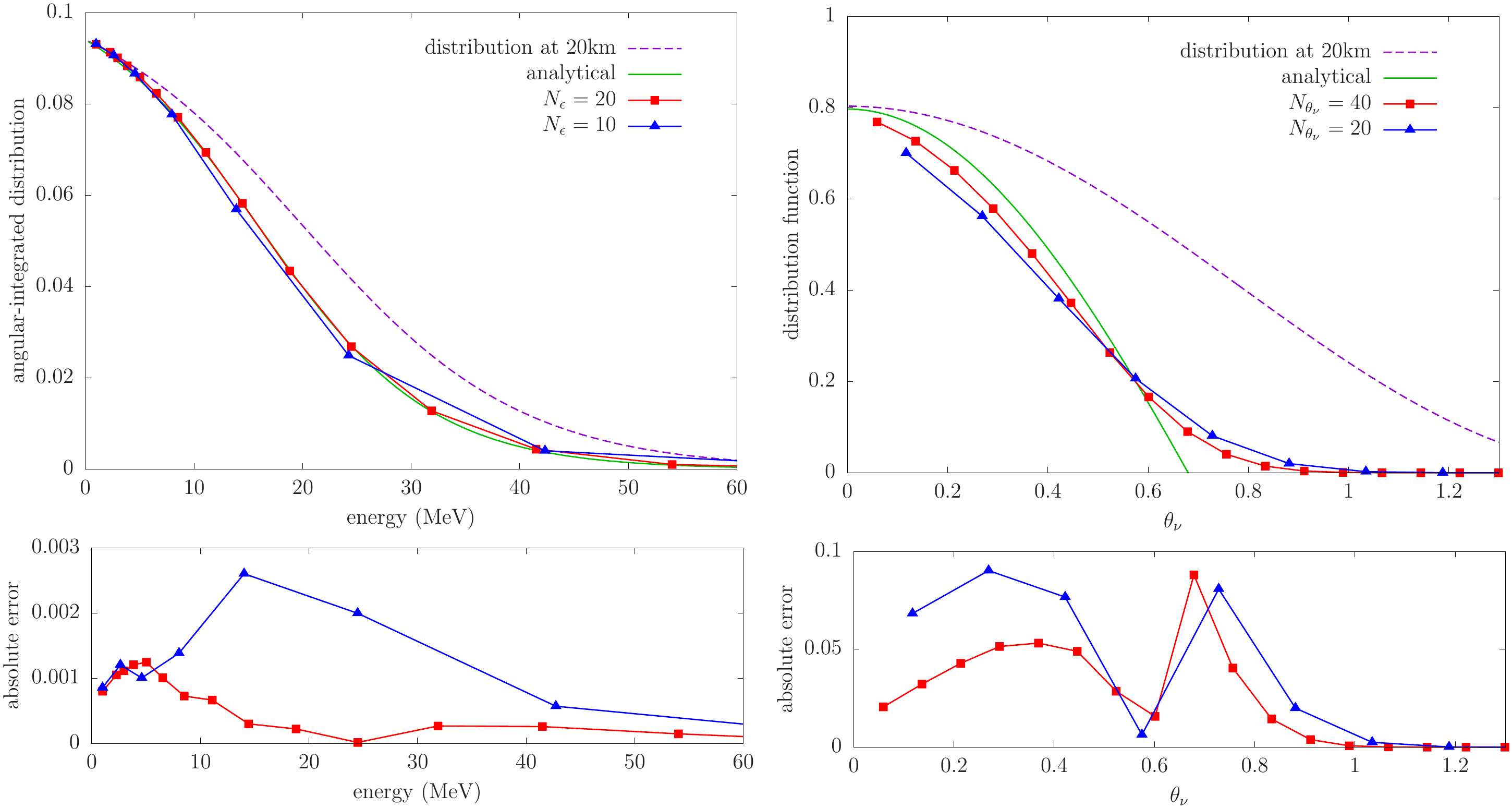}
  \caption{The neutrino distributions in energy and momentum angle at $40\,{\rm km}$. The top left panel shows the energy distribution of angular-integrated distribution function, and the top right panel shows the angular distribution of neutrinos with energy $\epsilon=5\,{\rm MeV}$. The green curves indicates analytical values, where blue and red curves indicate the numerical results for low and high resolution, respectively. The distribution at $20\,{\rm km}$ (source position) are shown with purple dotted curves for comparison. The bottom panels show the absolute errors for energy and angular distribution.}
  \label{Fig:fermiadv} 
\end{figure*}

Figure \ref{Fig:fermiadv} shows the angular-integrated energy distribution (top left panel), and angular distribution for the neutrino energy of $\epsilon=5\,{\rm MeV}$ (top right panel) at $r=40\,{\rm km}$. The dashed lines are the distributions at the source position while the solid lines give the analytic solutions (green) and the numerical results (blue for the lower resolution and red for the higher resolution). The bottom panels present the absolute values of errors. As can be seen, the energy spectrum is well reproduced already with 10 energy-grid points, which is actually smaller than the standard number employed in our recent CCSN simulations, and the numerical result gets even closer to the analytic solution with 20 energy-grid points. Such a converging feature is also seen for the angular distributions. By comparing the right panel of \ref{Fig:fermiadv} (result for smooth distribution) and figure \ref{Fig:tophat} (result for single-angular bin test \ref{sec:schwarzangadv}), it is intuitively understandable that numerical diffusion is much smaller for the extended distribution assumed here.

The results of the above test for the smooth distribution suggest that 20 energy bins employed in our CCSN simulations are large enough whereas 10 angular bins in momentum space are not sufficient at large radii, where the angular distribution becomes forward-peaked as assumed in this test. This is actually a well-known problem and is consistent with the previous investigation by \citet{Richers2017}. We note, however, that the number of these mesh points can be increased by a factor of 2 or more (depending on which number is increased) when the latest Japanese flagship supercomputer Fugaku is available soon, which is roughly $\sim40$ times faster than K supercomputer which we used for the SN simulations so far.

\subsection{Axisymmetric Tests}
\label{sec:radthe}
Our code is multi-dimensional in space. Here we test our code's capability to deal with the angular advection in space by calculating again the non-radial streaming of neutrinos in the Schwarzschild spacetime with this $\theta$-advection explicitly taken into account. 

We hence run the code in 2D under axisymmetry in this section. We compute the distribution function of neutrino on the $\phi$-constant meridional plane. The initial condition is as follows: we set $f=1$ for a single cell at $r=30\,{\rm km}$ and $\theta=\pi/2$ with $\theta_\nu=\pi/2$ and $\phi_\nu=\pi$ (i.e., moving in the positive z direction) in momentum space. Note that the neutrinos move along a geodesic curve also in this case that is essentially the same as the one in the previous test. We again switch off the energy advection in this test. Also, the $\phi_\nu$ advection is switched off, since neutrinos following the geodesic do not advect in that direction, in the current setting.

The radial mesh is the same as in section \ref{sec:1Dtest}, with $N_r = 128$. In addition to this, we deploy the $\theta$-mesh that has $N_\theta = 128$ bins, covering the range $\theta\in[0,\pi]$. As for the angular mesh in momentum space, we adopt $N_{\theta_\nu} = 10$, $20$, and $40$ to see the resolution dependence; the number of $\phi_\nu$-mesh points fixed to $N_{\phi_\nu} = 10$.

\begin{figure*}[t]
\begin{center}
  \includegraphics[width=18cm]{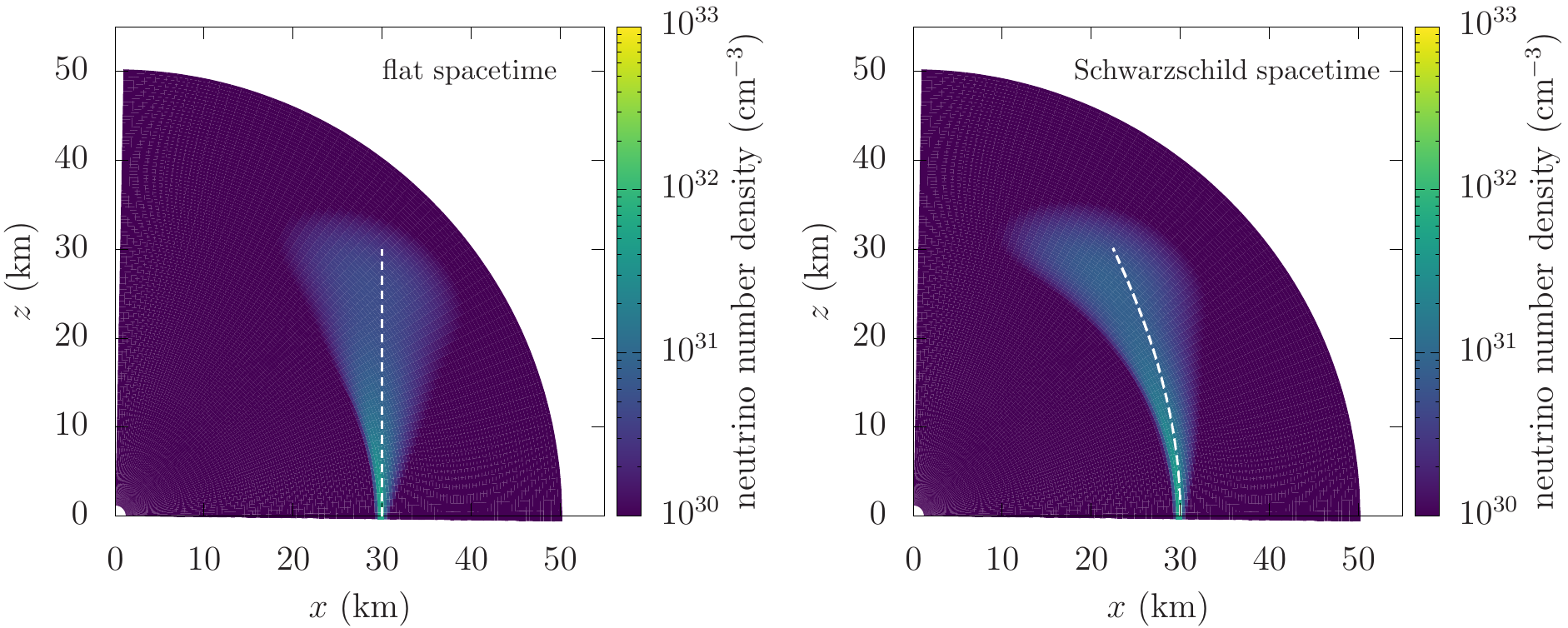}
  \caption{Neutrino number densities on the meridional plane. The horizontal and vertical axes are $x=r\mathrm{sin}\theta$ and $y=r\mathrm{cos}\theta$, respectively.)} The white dashed lines depict the geodesic curves drawn from the source to the position that the massless particles reach at the time of the snapshot. The left panel shows the result for the flat spacetime at time $t=1\times10^{-5}\,{\rm s}$, and the right panel is for the Schwarzschild spacetime at time $t=1.3\times10^{-5}\,{\rm s}$.
  \label{Fig:radthe}
  \end{center}  
\end{figure*}

Figure \ref{Fig:radthe} shows the neutrino number density, i.e., the distribution function integrated over the momentum space, on the meridional plane for $N_{\theta_\nu}=40$. The white dashed curve depicts the geodesic curve drawn from the source to the point that the massless particles reach at the time of the snapshot. The left panel is the result for the flat spacetime at the coordinate time $t=1\times10^{-4}\,{\rm s}$, where the geodesic is the straight line parallel to the $z$ axis. The right panel is for the Schwarzschild spacetime at $t=1.3\times10^{-4}\,{\rm s}$. It is apparent that the geodesic is deflected by gravity in this case. In both cases, the numerically obtained distributions are consistent with the analytical curves. On the other hand, the broadening of the beam is also evident. Just as in the previous tests, this is partly because the the beam has a finite width from the beginning and partly because there are numerical diffusions. 

We now quantify the numerical diffusion. This time we look at the neutrino propagation speed, which should be the speed of light but the diffusion will affect it. For this purpose we first evaluate the number densities on the geodesic as a function of $\theta$ by linearly interpolating the values on the neighboring radial cells as follows:
\begin{equation}
\label{eq:interpol}
N_\mathrm{int}(\theta) = \frac{
(r_{n} - r(\theta))N_{n-1}(\theta)
+ (r(\theta) - r_{n-1})N_n(\theta)}
{r_n-r_{n-1}},
\end{equation}
where $N_n(\theta)$ is the neutrino number density on the $n$-th radial mesh point for the given $\theta$, and $r(\theta)$ is the radial coordinate of the geodesic curve at the same angle $\theta$. We parametrize the geodesic not with $\theta$ but with the ``light-traveling distance'' defined as
\begin{equation}
\label{eq:schwarzdistance}
\displaystyle \int{\sqrt{g_{00}^{-1}(g_{11}dr^2+g_{22}d\theta^2)}},
\end{equation}
where $g_{\mu\nu}$ is the spacetime metric and the integration runs from the source to a point on the geodesic curve. This quantity has a simple physical interpretation: the light speed times the coordinate time it takes a massless particle to reach the point.

\begin{figure}[t]
  \includegraphics[width=8cm]{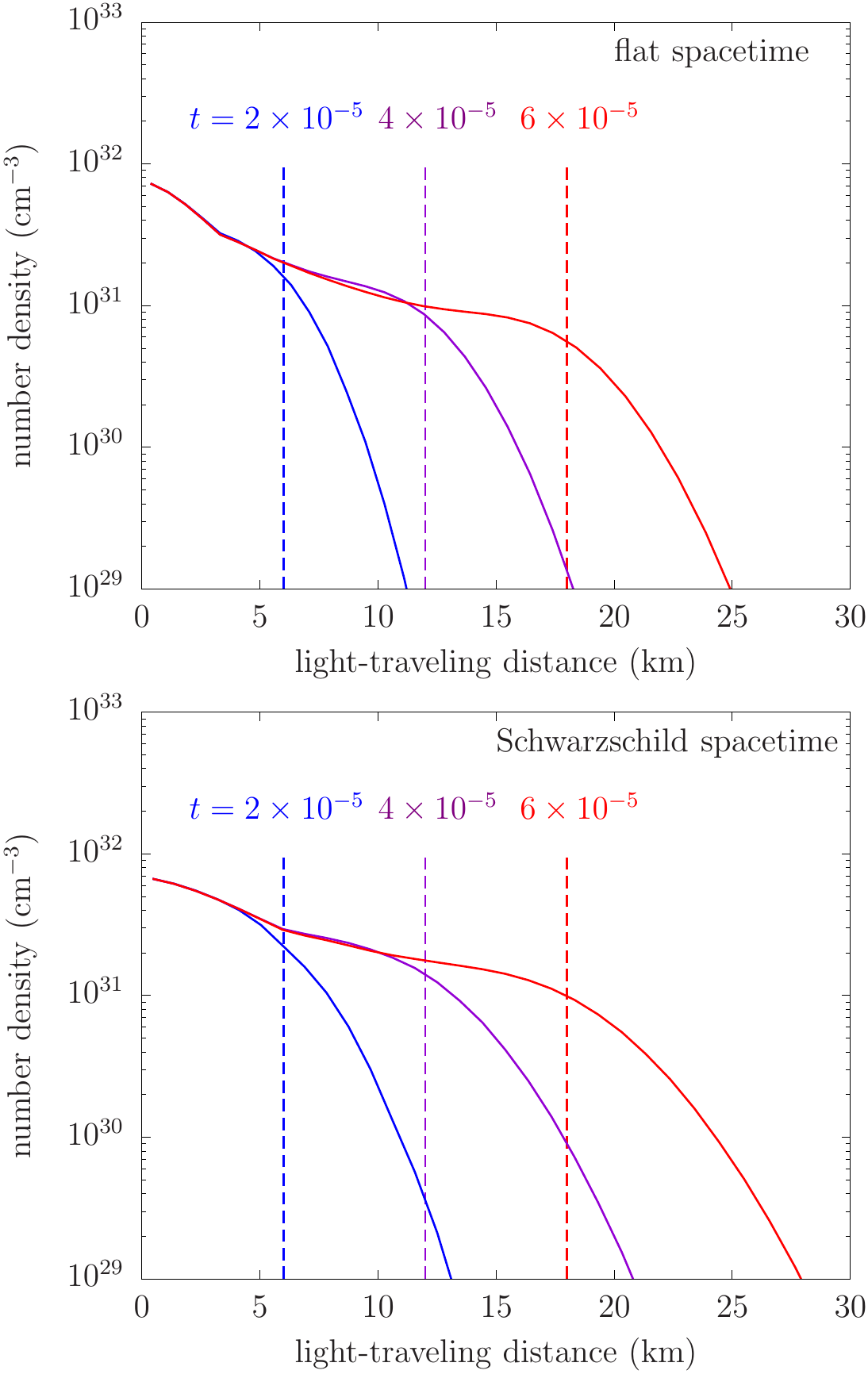}
  \caption{The profiles of the neutrino number density along the geodesic curve at different coordinate times. The upper and lower panels show the results for the flat and Schwarzschild spacetimes, respectively. The colors denote the times: the blue, purple, and red correspond to $t=2\times10^{-5}$, $4\times10^{-5}$, and $6\times10^{-5}\,{\rm s}$, respectively; the dashed lines indicate the exact positions of the front edge of the geodesic curves at these times.}
  \label{Fig:numgeo}
\end{figure}

Figure \ref{Fig:numgeo} shows the number density profile for different time steps along the geodesic obtained with equation \ref{eq:interpol} and parametrized with the light-traveling distance in equation \ref{eq:schwarzdistance}. The upper and lower panels show the results for the flat and Schwarzschild spacetimes, respectively. The dashed lines indicate the exact results. Although the number density is not constant owing to the beam broadening and there are some superluminal diffusions, the distribution declines rapidly ahead of the exact position and we may hence say that the propagation velocity is roughly consistent with the speed of light.

We finally study the resolution dependence, defining the following error function to quantitatively estimate the numerical diffusion:
\begin{equation}
\label{eq:rtherr}
E_{r\theta}(\theta)\equiv\frac{
\sum_{m=1}^{N_r}N_n(\theta)(r_n - r(\theta))^2dr_n}
{(r(\theta))^2\sum_{n=1}^{N_r}N(r_n,\theta)dr_n},
\end{equation}
where $r_n$ and $dr_n$ are the radial coordinate at the center and the width of the $n$-th radial cell, respectively, and $r(\theta)$ is the radius of the point on the geodesic curve at $\theta$.

\begin{figure*}[t]
\begin{center}
  \includegraphics[width=18cm]{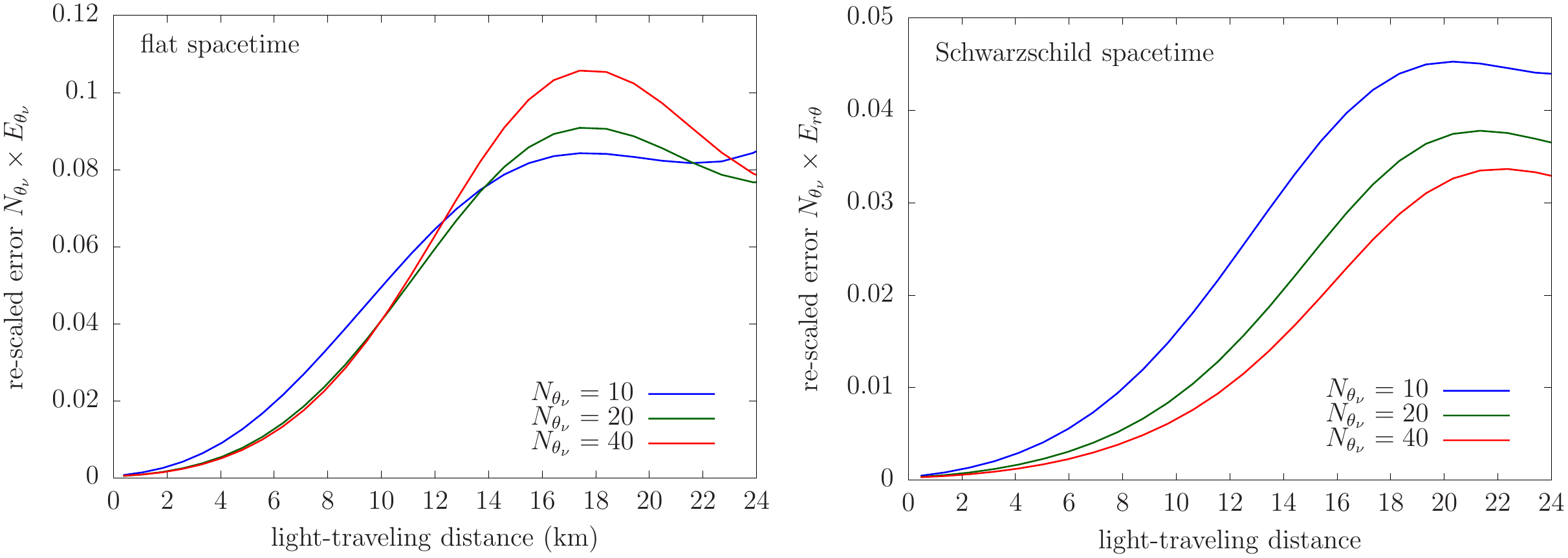}
  \caption{The error function in equation (\ref{eq:rtherr}) re-scaled with the number of angular mesh points $N_{\theta_\nu}$. The left and right panels show the results for the flat and Schwarzschild spacetimes, respectively. The blue, green, and red curves correspond to the errors for the $N_{\theta_\nu} = 10$, $20$, and $40$, respectively.}
  \label{Fig:rtherr}
  \end{center}
\end{figure*}

Figure \ref{Fig:rtherr} shows the re-scaled error function, which is defined as the error function in equation (\ref{eq:rtherr}) multiplied by $N_{\theta_\nu}$; although the equation (\ref{eq:rtherr}) is a function of $\theta$, we employ the light-traveling distance to parametrize the geodesic in this figure. The left and right panels again present the results for the flat and Schwarzschild spacetimes, respectively. It is recognized that both cases roughly show the first-order convergence.

\section{Advection Tests in the Kerr Spacetime}
\label{sec:kerrtest}
We now move on to the advection tests in the Kerr spacetime. They are intended to demonstrate the applicability of our code to rotating BH spacetimes. We ignore all neutrino reactions again in this section.

We employ the Kerr-Schild coordinates:
\begin{eqnarray}
ds^2 && =
- \left(1-\frac{2GMr}{c^2\Sigma}\right)c^2dt^2
\nonumber \\
&& + \frac{4GMr}{c^2\Sigma}\left[\frac{\Sigma}{r^2+a^2}dr-a\mathrm{sin}^2\theta d\phi\right]cdt \nonumber \\
&& + \left(\frac{\Sigma}{r^2+a^2}+\frac{2GMr\Sigma}{c^2(r^2+a^2)^2}\right)dr^2
\nonumber \\
&&
+ \Sigma d\theta^2
+ \frac{\Xi}{\Sigma}\mathrm{sin}^2\theta d\phi^2
- \frac{4GMra}{c^2(r^2+a^2)}\mathrm{sin}^2\theta dr d\phi, \nonumber \\
\end{eqnarray}
where $a$ is the BH spin parameter, and
\begin{eqnarray}
&& \Sigma = r^2+a^2\mathrm{cos}^2\theta,\nonumber \\
&& \Xi = (r^2+a^2)\Sigma + 2GMa^2r\mathrm{sin}^2\theta/c^2.
\end{eqnarray}
Note that there is no (apparent) singularity at the event horizon in these coordinates. Throughout this section, we choose $M=5M_\odot$ and $a=0.5GM/c^2$.
The tetrad in equation (\ref{eq:tetrad}) is explicitly given as follows:
\begin{eqnarray}
&&
e_{(0)}^\mu
= \frac{1}{\alpha}
\left[
\left(\frac{\partial}{\partial t}\right)^\mu
- \beta^r\left(\frac{\partial}{\partial r}\right)^\mu
- \beta^\phi\left(\frac{\partial}{\partial \phi}\right)^\mu
\right],
\nonumber \\
&&
e_{(1)}^\mu
= \frac{1}{\sqrt{\gamma_{rr}}}
\left(\frac{\partial}{\partial r}\right)^\mu,
\nonumber \\
&&
e_{(2)}^\mu
= \frac{1}{\sqrt{\Sigma}}
\left(\frac{\partial}{\partial \theta}\right)^\mu,
\nonumber \\
&&
e_{(3)}^\mu
= \frac{\gamma^{r\phi}}{\sqrt{\gamma^{\phi\phi}}}
\left(\frac{\partial}{\partial r}\right)^\mu
+ \frac{1}{\sqrt{\gamma^{\phi\phi}}}
\left(\frac{\partial}{\partial \phi}\right)^\mu,
\end{eqnarray}
where the lapse function, shift vector, and inverse of the spatial metric are written as
\begin{eqnarray}
&&
\alpha=\sqrt{\frac{c^2\Sigma}{c^2\Sigma+2GMr}},
\nonumber \\
&&
\beta^r=\frac{2GMr}{c^2\Sigma+2GMr},
\nonumber \\
&&
\beta^\phi=-\frac{2GMar}{(c^2\Sigma+2GMr)(r^2+a^2)},
\nonumber \\
&&
\gamma^{rr}=\frac{c^2\Xi}{\Sigma(c^2\Sigma+2GMr)},
\nonumber \\
&&
\gamma^{r\phi}=\frac{2GMar}{(c^2\Sigma+2GMr)(r^2+a^2)},
\nonumber \\
&&
\gamma^{\phi\phi}=\frac{\Sigma(r^2+2GMr/c^2+a^2)}{(r^2+a^2)^2(\Sigma+2GMr/c^2)\mathrm{sin}^2\theta},
\end{eqnarray}
respectively. As mentioned in \citet{Shibata2014}, there is no coordinate singularity if we use this tetrad in the Kerr-Schild coordinates. 

As in section \ref{sec:schwarzangadv}, we perform the advection test by placing a point source in the Kerr spacetime. We treat the neutrino propagation only on the equatorial plane ($\theta=\pi/2$) as our main concern here is the dragging of inertial frame. We fix $f=1$ for a single angular bin at the position of the point source and set $f=0$ otherwise initially. The energy advection is turned off again in this case. We set the initial direction of neutrino momentum to $\phi_\nu=3\pi/2$, i.e., the retrograde direction with respect to the BH spin, to maximize the frame-dragging effect. We further assume $p^r=0$ to distinguish the geodesic curves outside the photon sphere from those inside clearly. There appears a bit complication then, because this does not correspond to $\theta_\nu=\pi/2$ for our choice of tetrad in the Kerr spacetime. That happens because the coordinate basis is not orthogonal. We need to find the value of $\theta_\nu$ by solving the following equation:
\begin{equation}
p^r =
\epsilon(e^1_{(0)}
+ e^1_{(1)}\mathrm{cos}\theta_\nu
- e^1_{(3)}\mathrm{sin}\theta_\nu) = 0.
\end{equation}
This yields, for example, $\theta_\nu=0.9668$ for $r=28\,{\rm km}$ and $\theta_\nu=0.7695$ for $r=22\,{\rm km}$.

Throughout this test, the radial mesh is the same as in the previous tests, deploying $N_r=128$ grid points. We vary the number of angular mesh points as $N_{\theta_\nu} = 30$, $40$, and $60$ to see the resolution dependence.

\begin{figure*}[t]
\begin{center}
  \includegraphics[width=18cm]{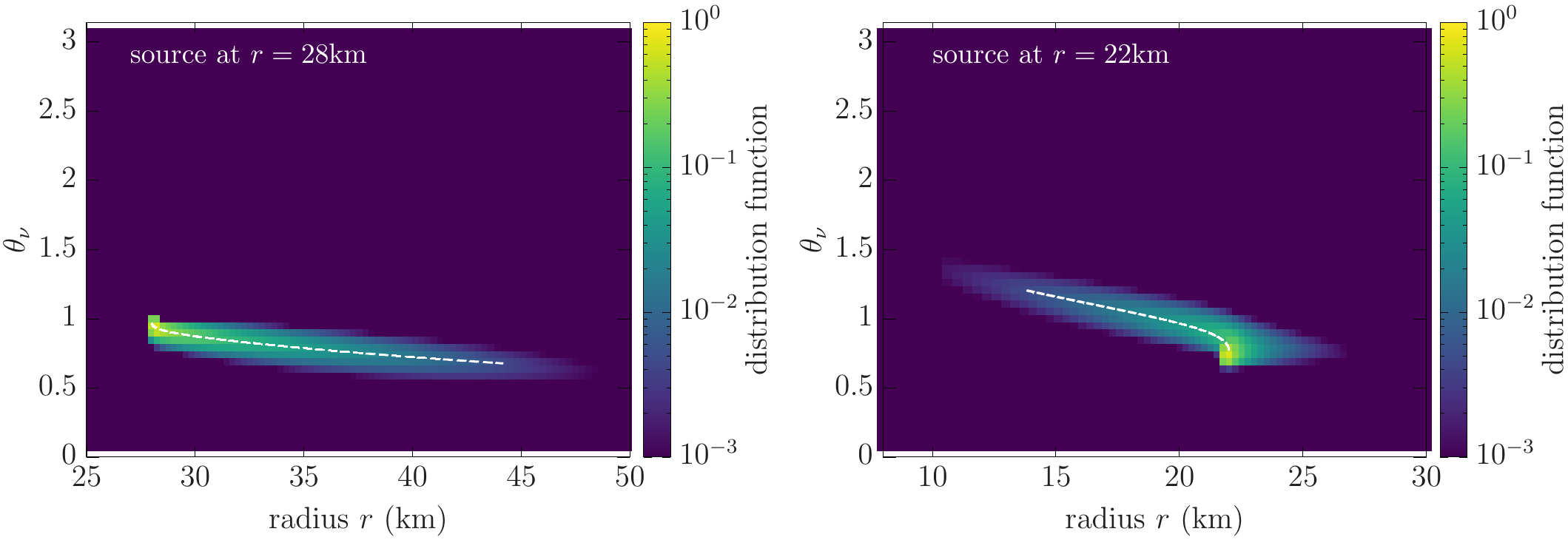}
  \caption{Angular neutrino distributions along the geodesic curve on the equatorial plane as a function of the radius $r$ in the Kerr spacetime. The left panel is the result at time $t=4\times10^{-4}\,{\rm s}$, with the source located at $r=28\,{\rm km}$ outside the photon sphere. The right panel is the result at time $t=1.5\times10^{-4}\,{\rm s}$, with the source located at $r=22\,{\rm km}$ inside the photon sphere. The white dashed curves are the reference geodesic curves, drawn from the source to the radius that the massless particles reach at the time of the snapshot.}
  \label{Fig:kerr_out_in}
  \end{center}
\end{figure*}

Figure \ref{Fig:kerr_out_in} shows the angular distribution in momentum space as a function of the radius $r$ for $N_{\theta_\nu}=60$. The white dashed curve is the reference geodesic curve calculated in appendix \ref{sec:kerr}, drawn from the source to the radius that the massless particles reach at the time of the snapshot. The left panel shows the result at the coordinate time $t=4\times10^{-4}\,{\rm s}$ with the source located at $r=28\,{\rm km}$. In this case, the source is sitting outside the photon sphere ($r=26\,{\rm km}$) and the neutrinos propagate radially outward. The right panel shows the result at $t=1.5\times10^{-4}\,{\rm s}$ with the source located at $r=22\,{\rm km}$, i.e., inside the photon sphere, and the geodesic goes radially inward. The broadening of the beam is again apparent, which is inevitable as the mesh size is finite, and in the latter case, in particular, there are a fraction of neutrinos going radially outward due to the numerical diffusion. Nevertheless, most of neutrinos propagate consistently with the geodesic curves in both cases. We may hence claim that our code can handle the advection also in the rotating spacetime.

\begin{figure}[t]
  \includegraphics[width=8cm]{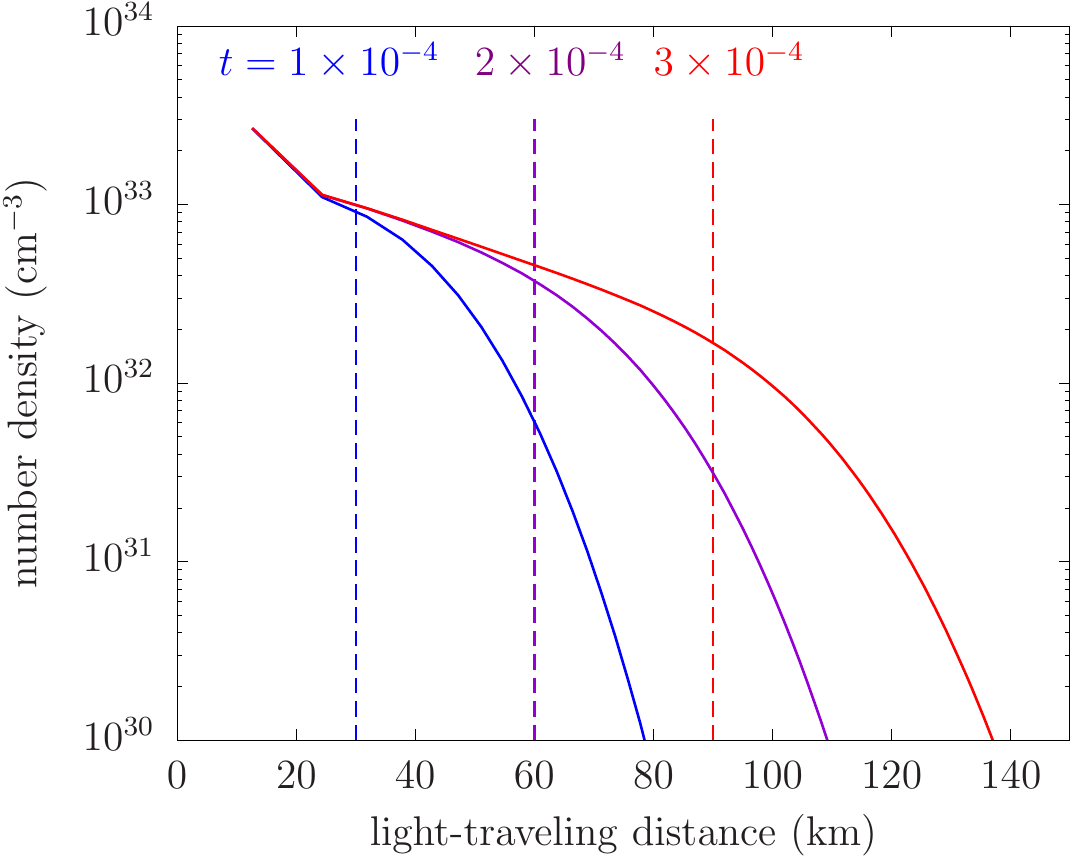}
  \caption{The profiles of neutrino number density along a geodesic curve in the Kerr metric at three different coordinate times: the blue, purple, and red colors correspond to $t=1\times10^{-4}$, $2\times10^{-4}$ and $3\times10^{-4}\,{\rm s}$, respectively. The dashed lines show the exact positions that the massless particles reach at these times.}
  \label{Fig:numgeokerr}
\end{figure}

We look at the neutrino propagation speed. Since our computation does not take the $\phi$ advection into account directly thanks to the axisymmetry, we inspect the radial propagation. The geodesic is again parametrized by the light-traveling distance defined in the current case as
\begin{eqnarray}
\displaystyle
&&
\int(g_{00})^{-1}[-(g_{01}dr+g_{03}d\phi)
\nonumber \\
&&
-\sqrt{(g_{01}dr+g_{03}d\phi)^2
-g_{00}(g_{11}dr^2+g_{33}d\phi^2+2g_{13}drd\phi)}\,].\nonumber \\
&&
\end{eqnarray}
It has the same physical interpretation as in the Schwarzschild spacetime (equation (\ref{eq:schwarzdistance})). Figure \ref{Fig:numgeokerr} shows the profiles of the number density along the geodesic as a function of the light-traveling distance. Although the the results are consistent with the propagation at the speed of light, the numerical diffusion is more remarkable than in the flat or Schwarzschild spacetime. This is due to the slower radial propagation of neutrinos in the current case, which is in turn caused by the frame-dragging in the rotating spacetime. Recall that we are considering the retrograde advection here.

\begin{figure}[t]
  \includegraphics[width=8cm]{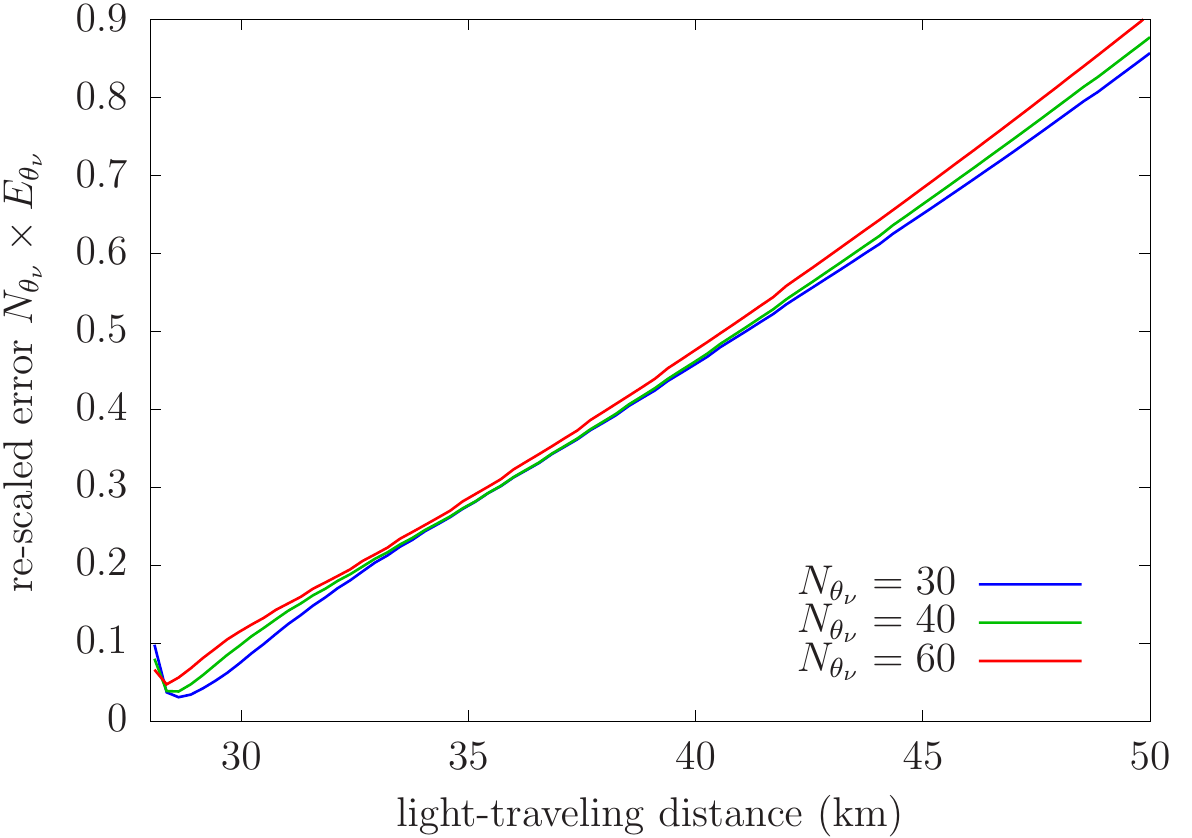}
  \caption{Radial profiles of the error function re-scaled with the number of angular mesh points $N_{\theta_\nu}$. The blue, green, and red colors are the results for the different mesh numbers $N_{\theta_\nu} = 30$, $40$, and $60$, respectively.}
  \label{Fig:kerrerr}
\end{figure}

Using the error function given in equation (\ref{eq:angerr}), we show in figure \ref{Fig:kerrerr} the re-scaled error function for the advection test in the Kerr spacetime. It is apparent that the error scales linearly with the number of angular mesh points just as in the previous tests in the flat or Schwarzschild spacetime.

\section{Tests Including Collision Terms}
\label{sec:collision}
We have so far conducted various verification tests in the free-streaming limit alone. This is because the computation of advection is the most challenging numerically for mesh-based codes like ours. In the actual astrophysical simulations, collision terms of the Boltzmann equation are also important. In this section, we perform a test that takes into account interactions of neutrinos with matter.
 
We fix the background matter distribution and compute with our new code the neutrino distribution that approaches a steady state. Since the collision terms are the focus of this test, we assume spherical symmetry again. As a reference, we employ the results obtained with another spherically symmetric, general relativistic Boltzmann solver (hereafter 1D GR code) developed by \citet{Sumiyoshi2005}. We run these two Boltzmann solvers for the same (fixed) matter distribution until the steady states are achieved. As for the matter distribution, we pick up a snapshot at 100 ms after bounce in the CCSN simulation by \citet{Sumiyoshi2005} with the $15\,M_\odot$ model of \citet{Woosley1995}. It allows us to test the neutrino propagation in a wide range of mean free path. The major neutrino-matter reactions are the same for both simulations, the rates of which are based on \citet{Bruenn1985}. The explicit expressions of the collision terms are identical to those described in \citet{Sumiyoshi2012}. The chemical potential profile is also common because it affects the reaction rates: it is calculated from Shen's equation of state \citep{Shen1998}. In the 1D GR code, the hydrodynamics variables as well as the spacetime metric are the functions of the enclosed mass as it is a Lagrangian code. However, the fluid velocity is fixed to zero. They are transformed into the functions of the radius when they are implemented in the new code. 

The numbers of the radial mesh points are the same for the two codes: the radial mesh has $N_r = 256$ grid points, which covers the range $r\in [0,10^4]\,{\rm km}$. As for the momentum space, we adopt $N_{\theta_\nu} = 6$ and $N_\epsilon = 14$. The latter mesh covers $\epsilon\in [0,300]\,{\rm MeV}$. The reader may be worried that the resolution employed here are relatively lower compared to the advection tests in sections \ref{sec:schwarzadv} and \ref{sec:kerrtest}, or recent core collapse simulations. This is no problem, since our purpose is the regression test, i.e. to confirm that the collision terms work properly.

\begin{figure}[t]
  \includegraphics[width=8.0cm]{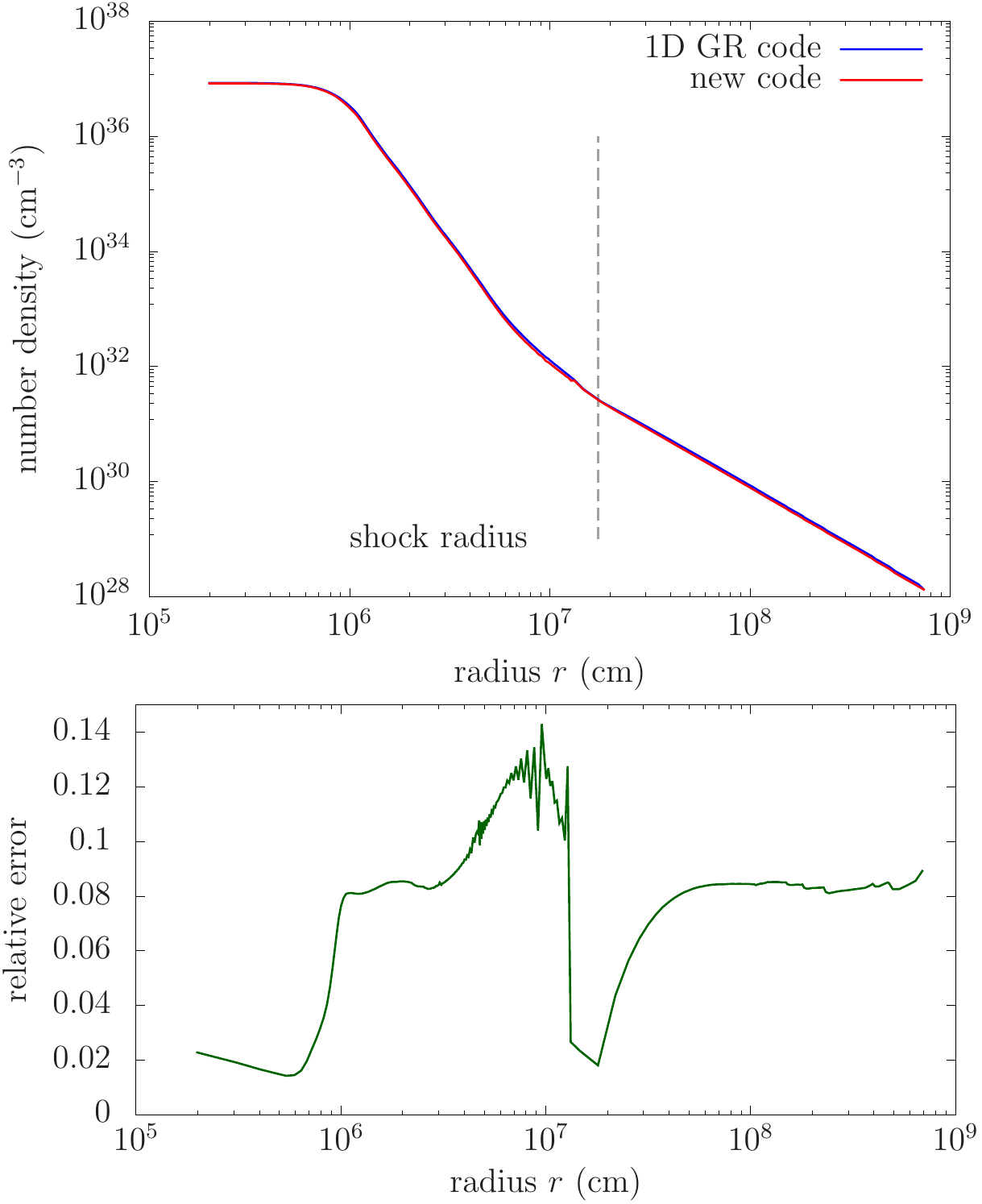}
  \caption{Radial profiles of the electron-type neutrino number density for the 1D code and the new code (upper panel) and their relative errors (lower panel). The blue and red curves in the upper panel represent the results for the 1D GR and new codes, respectively.}
  \label{Fig:1DGRnum}
\end{figure}

Figure \ref{Fig:1DGRnum} shows the radial profiles of the electron-type neutrino number density for our new code and the 1D GR code (upper panel), as well as the relative difference with respect to the latter result (lower panel). The shock wave is located at $r=174 \,{\rm km}$, at which a small bump can be seen. It is apparent that they are in a reasonable agreement with the overall relative error of $\sim10\%$ ($\sim14\%$ at the maximum). It is particularly small $\sim1\%$ near the center ($r \lesssim 10 \,{\rm km}$), where the neutrino distribution is close to that of thermal equilibrium.

\begin{figure}[t]
  \includegraphics[width=8.0cm]{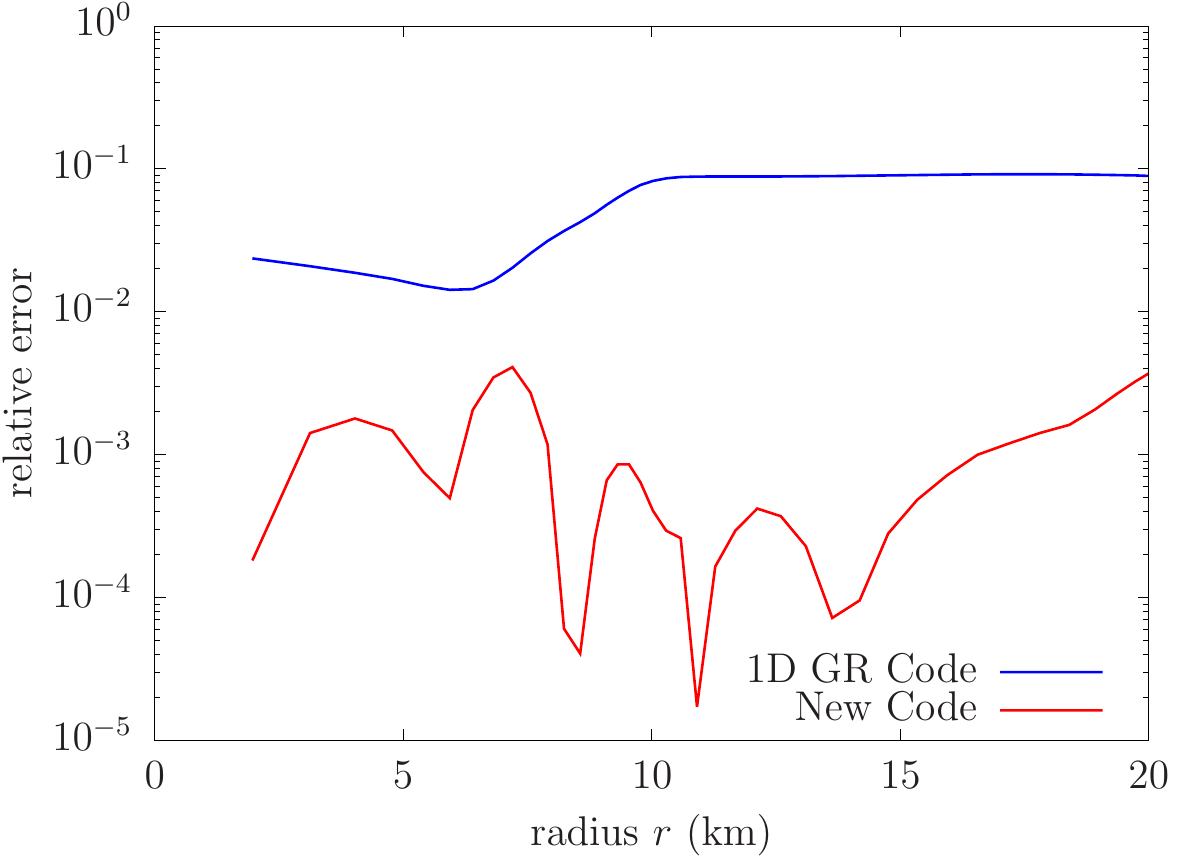}
  \caption{Relative deviations of the simulated neutrino number densities from the equilibrium values. The blue and red curves represent the results for the 1D GR and new codes, respectively.}
  \label{Fig:FDerrN}
\end{figure}

It should be noted that the above comparison does not tell which code is more accurate. To get some hints, we compare them with the equilibrium number densities, which are obtained by the energy integration of the Fermi--Dirac distribution. Figure \ref{Fig:FDerrN} shows the relative deviation of the neutrino number density obtained with each of the two codes from the equilibrium density. As is clear, the difference is much smaller for the new code in the very optically thick regime. It is thought to come from the different way of evaluating the equilibrium distribution for the calculation of the reaction kernels. The 1D GR code simply calculates the equilibrium distribution by the value of energy at the center of the cell, whereas the new code derives the equilibrium value as the average over a energy bin, by dividing it into subgrid. Incidentally, neither result is very accurate at large radii because of the rather coarse angular resolutions employed.

Finally, we give some results of the comparison with the M1 closure, the currently most popular approximation in the literature. In the neutrino transport with the M1 closure, second angular moment $P^{ij}$ of the distribution function in momentum space is given as the interpolation of the optically thin and thick limits:
\begin{equation}
P^{ij}(\epsilon) = \frac{3\chi-1}{2}P^{ij}_\mathrm{thin}(\epsilon) + \frac{3(1-\chi)}{2}P^{ij}_\mathrm{thick}(\epsilon),
\end{equation}
where the optically thin and thick limits are given as
\begin{equation}
P^{ij}_\mathrm{thin}(\epsilon) = E(\epsilon)\frac{F^i(\epsilon)F^j(\epsilon)}{\gamma_{kl}F^k(\epsilon)F^l(\epsilon)},\quad P^{ij}_\mathrm{thick}(\epsilon) = \frac{\delta^{ij}}{3},
\end{equation}
respectively, where $E(\epsilon)$ and $F^i(\epsilon)$ are the energy density and flux, respectively. Note that we only treat the case where fluid velocity is zero throughout this discussion. The Eddington factor $\chi$ is given, for example, as \citep[][]{Levermore1984}
\begin{equation}
\chi = \frac{3+4\bar F^2}{5+2\sqrt{4-3\bar F^2}},
\end{equation}
where the flux factor $\bar F$ may be calculated as \citep{Shibata2011}
\begin{equation}
\bar F = \frac{\sqrt{\gamma_{ij}F^i(\epsilon)F^j(\epsilon)}}{E(\epsilon)}.
\end{equation}
In the Boltzmann transport, the second moment can be directly calculated from the distribution function, and in spherical symmetry, the Eddington factor is $\chi=k^{rr}$, where $k^{ij}(\epsilon)=P^{ij}(\epsilon)/E(\epsilon)$ is the Eddington tensor.

\begin{figure}[t]
\includegraphics[width=8cm]{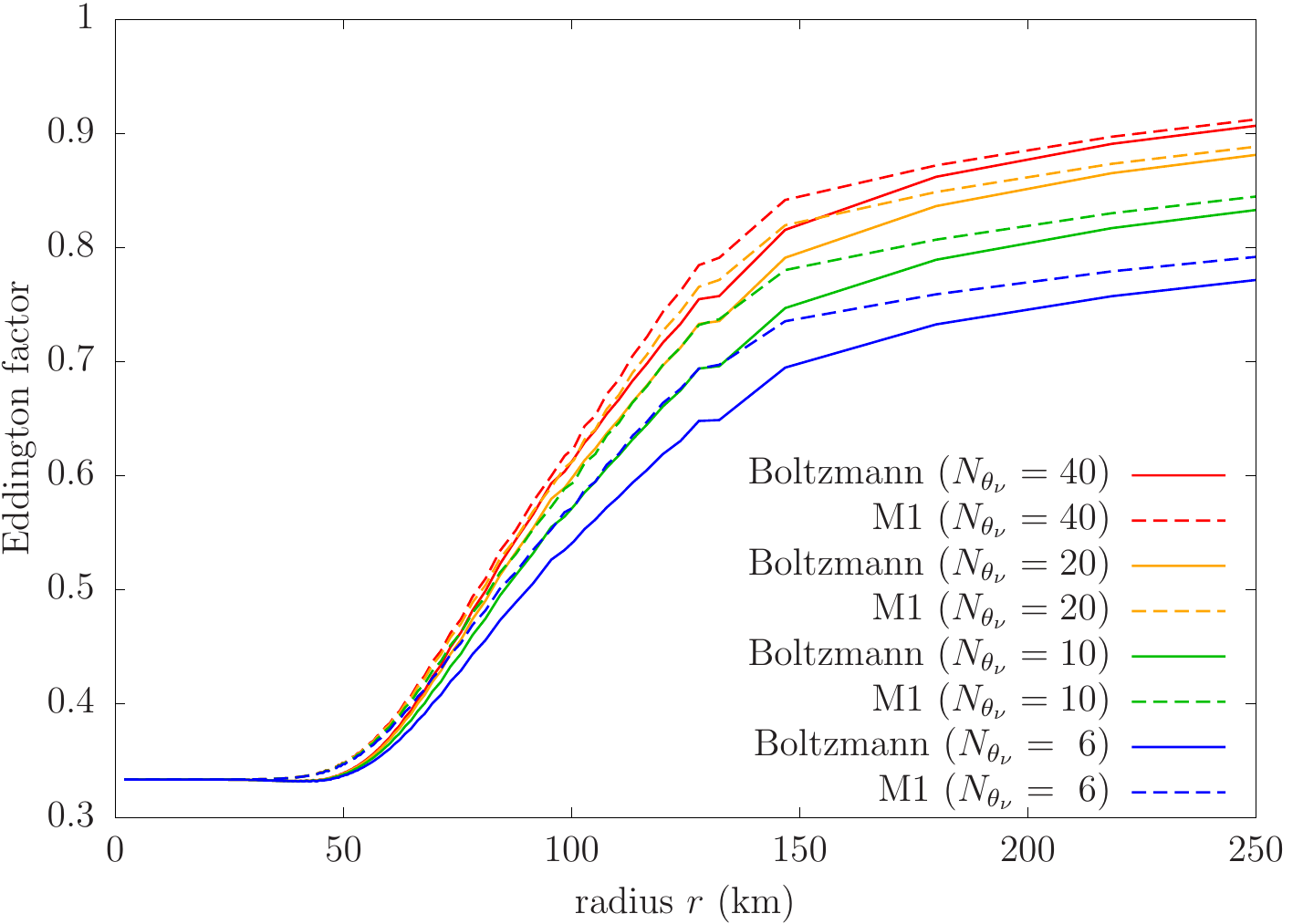}
  \caption{Radial profiles of the Eddington factor. The solid curves and dotted curves correspond to Boltzmann and M1, respectively. Different colors indicate the numbers of angular mesh points: blue, green, orange and red curves are for $N_{\theta_\nu}=6$, $10$, $40$, respectively.}
  \label{Fig:eddfac} 
\end{figure}

Figure \ref{Fig:eddfac} shows the radial profiles of the Eddington factors obtained directly by the Boltzmann solver (solid lines) and approximately with the M1 closure relation (dashed lines). It is observed that the M1 closure tends to give larger values than the Boltzmann solver; the latter with lower angular resolutions tends to underestimate the Eddington factor particularly at large radii where the neutrino angular distribution becomes forward-peaked; the M1 closure fails, on the other hand, in the semi-transparent region $\sim140\,{\rm km}$. All these results are quantitatively consistent with those found in our previous studies in the Minkowski spacetime \citep{Harada2019}. It is worth mentioning that, however, that this is not a true comparison between the Boltzmann transport and the two-moment transport with the M1 closure. In this analysis of the M1 closure approximation, the Eddington factor is calculated from the energy density and the flux obtained by the Boltzmann solver. In actual two-moment transport, they should be computed on their own and errors may be accumulated with time and could be larger than suggested in figure \ref{Fig:eddfac}. In the mean time, 10 grid points in $\theta_{\nu}$ is not sufficient in our Boltzmann solver and should be doubled or more, which has actually come in sight.)

\section{Summary and Discussions}
\label{sec:sum&dis}
We have developed a numerical code to solve the multidimensional Boltzmann equation for neutrino transfer in full general relativity, employing the discrete ordinate method. We have performed a series of tests to confirm that our code can handle the general relativistic neutrino transport. We first conducted the tests of free streaming in the Schwarzschild and the Kerr spacetimes. We computed the propagation of neutrinos from a source located either inside or outside the photon sphere in these spacetimes. We confirmed that the results are consistent with the reference geodesic curves. We also found that the deviation from the reference is inversely proportional to the number of mesh points either in energy or zenith angle, implying that the numerical accuracy of the energy or angular advection is of the first order as expected from our finite-difference schemes adopted for these terms.

Next we ran the code with the collision terms included for a fixed matter distribution extracted from our realistic 1D simulation of CCSN. We demonstrated that the result is consistent with that obtained with another 1D GR code developed by \citet{Sumiyoshi2005}. We also found that our new code can reproduce the thermal equilibrium distribution in the optically thick region more accurately than the 1D GR code.

In this paper we ignore the evolutions of matter configurations and spacetime entirely. Having completed the development of the Boltzmann solver, our next task is to combine it with the general relativistic hydrodynamics code and the solver of the Einstein equations, both of which have already been developed separately. The latter, in particular, is based on the Baumgarte-Shapiro-Shibata-Nakamura formalism \citep{Shibata1995, Baumgarte1998} formulated for the polar coordinates, employing the partially-implicit Runge Kutta method \citep{Baumgarte2014}. Their performance tests will be reported in elsewhere in the near future. Once completed we will apply the integrated code to various general relativistic phenomena.


\acknowledgments

Numerical computations were in part carried out on Cray XC50 at Center for Computational Astrophysics, National Astronomical Observatory of Japan, and the supercomputer SX-Aurora the Particle, Nuclear and Astrophysics Simulation Program (No. 2019-002, 2020-004) at the High Energy Accelerator Research Organization (KEK). This work is supported by Grant-in-Aid for Scientific Research (19K03837, 20H01905) and for Research Activity Start-up (19K23435) from Japan Society for the Promotion of Science (JSPS), and Grant-in-Aid for Scientific Research on Innovative areas "Gravitational wave physics and astronomy: Genesis" (17H06357, 17H06365) from the Ministry of Education, Culture, Sports, Science and Technology (MEXT), Japan. S. Y. is supported by Institute for Advanced Theoretical and Experimental Physics, and Waseda University and the Waseda University Grant for Special Research Projects (project number: 2020-C273). For providing high performance computing resources, Computing Research Center, KEK, JLDG on SINET of NII, Research Center for Nuclear Physics, Osaka University, Yukawa Institute of Theoretical Physics, Kyoto University, Nagoya University, Hokkaido University, and Information Technology Center, University of Tokyo are acknowledged. This work was supported by MEXT as "Program for Promoting Researches on the Supercomputer Fugaku" (Toward a unified view of the universe: from large scale structures to planets). This research used high-performance computing resources of the K-computer / the supercomputer Fugaku provided by RIKEN, the FX10 provided by Tokyo University, the FX100 provided by Nagoya University, the Grand Chariot provided by Hokkaido University, and Oakforest-PACS provided by JCAHPC through the HPCI System Research Project (Project ID: hp130025, 140211, 150225, 150262, 160071, 160211, 170031, 170230, 170304, 180111, 180179, 180239, 190100, 190160, 200102, 200124).

%





\appendix

\section{Evaluation of Advection Terms}
\label{sec:advfindif}
In this appendix, we describe how the advection terms are evaluated in our finite difference scheme. It is the general relativistic extension from the Newtonian counterpart given in \citet{Sumiyoshi2012}. Throughout this section, variables with the lowercase subscript $n$ (such as $r_n$) are defined at the cell centers whereas those with the uppercase subscript $N$ (such as $r_N$) are defined at the cell interfaces.

The evaluation of the advection terms in the discretized form is tricky when they are written in the conservative form. This is because there are some delicate cancellations among the terms that are a part of different advection terms. This may be understood if one considers a constant $f$ in the phase space. Then all advections should be vanishing. This is apparent in the Boltzmann equation written as in equation \ref{eq:boltz}, since all the derivatives of $f$ are zero. This is not so obvious, however, if the Boltzmann equation is cast into the conservative form, as in equation \ref{eq:conservBoltz}. In fact, the advection terms are not simply vanishing but are cancelled among them in nontrivial ways. Analytically these cancellations are with no problem; it is difficult to enforce in the finite-difference, however. If one treats them carelessly, neutrinos may appear from nowhere. The formulation described in \citet{Sumiyoshi2012} ensures the perfect cancellation of those terms for the flat spacetime. In the general relativistic case with an arbitrary metric, the formulation for the flat spacetime is not applicable, though. In this paper, we do not stick to the perfect cancellation and evaluate the advection terms are evaluated in a rather straightforward way as described below. As shown in the advection tests in sections \ref{sec:schwarzadv} and \ref{sec:kerrtest}, however, we found no problem in the general relativistic cases even if the cancellation is not exactly enforced. We will investigate this issue further and, if necessary, revise the general relativistic formulation in future works.

The numerical grid employed in this paper is based on the original flat version of the code, and the details are described in \citet{Sumiyoshi2012}. The original code employed $\mu\equiv\mathrm{cos}\theta$ and $\mu_\nu\equiv\mathrm{cos}\theta_\nu$ instead of $\theta$ and $\theta_\nu$, in order to simplify the advection terms and the six coordinate variables ($r$, $\mu$, $\phi$, $\epsilon$, $\mu_\nu$ and $\phi_\nu$) were discretized as described in the paper. In our general relativistic extension, on the other hand, the use of $\mu$ is not so useful, though. We hence use $\theta$ as an independent variable but give its values at the cell centers and interfaces according to $\theta_i = \mathrm{cos}^{-1}(\mu_i)$ and $\theta_I = \mathrm{cos}^{-1}(\mu_I)$, from the corresponding values of $\mu$.

The spatial advection terms ($r$, $\theta$ and $\phi$) are finite-differenced as
\begin{eqnarray}
&& 
\frac{1}{\sqrt{-g}}\frac{\partial}{\partial x^i}
\left[L^i\sqrt{-g}f\right]
\nonumber \\
\longrightarrow  
&&
\frac{1}{\sqrt{-g_n}dx^i_n}
\left[\sqrt{-g_N} L^i_N f_N
- \sqrt{-g_{N-1}} L^i_{N-1} f_{N-1}\right],
\nonumber \\
\end{eqnarray}
where $dx^i_n$ is the width of the $n$-th cell, and we defined $L^\mu \equiv e_{(0)}^\mu + \sum_{i=1}^{3}l_i e_i^\mu$ for notational simplicity. In the flat or the Schwarzschild case, for example, $L^1=\mathrm{cos}\theta_\nu$, $L^2=\mathrm{sin}\theta_\nu\mathrm{cos}\phi_\nu$, $L^3=\mathrm{sin}\theta_\nu\mathrm{sin}\phi_\nu$. The value of $L_Nf_N$ on the $N$-th cell interface ($L^i_Nf_N$) is evaluated as
\begin{equation}
L^i_N f_N = \frac{L^i_n - |L^i_n|}{2}[(1-\beta_N)f_n + \beta_N f_{n+1}] + \frac{L^i_{n+1} + |L^i_{n+1}|}{2}[\beta_N f_n + (1-\beta_N)f_{n+1}],
\end{equation}
where $\beta_N$ is introduced to smoothly switch from the upwind differencing in the free streaming limit ($\beta_N=1$) to the central differencing in the diffusion limit ($\beta_N=1/2$). We use the following expression of $\beta_N$ based on \citet{Mezzacappa1993}:
\begin{equation}
\label{eq:beta}
\beta_N = 1 - \frac{1}{2}\frac{\alpha_p dr_N\lambda_N}{1+\alpha_p dr_N\lambda_N}.
\end{equation}
where the mean free path on the cell interface is calculated as $\lambda_N=(\lambda_{n+1}+\lambda_n)/2$. The adjustable parameter $\alpha_p$ is set to be $100$, following \citet{Sumiyoshi2012}. In the free streaming case $\beta_N=1$, the upwind direction is accounted for by the signature of $L^i$.

The energy advection term is expressed as follows:
\begin{eqnarray}
&&  -\frac{\partial}{\partial(\epsilon^3/3)}
\left(\epsilon^3 f\omega_{(0)}\right)
\nonumber \\
\longrightarrow 
&&
-\frac{3}{\epsilon_N^3-\epsilon_{N-1}^3}
\left[
\epsilon_N^3 \omega_{(0)N} f_N
- \epsilon_{N-1}^3 \omega_{(0)N-1} f_{N-1} \right].
\end{eqnarray}
with
\begin{equation}
\omega_{(0)N} f_N = \frac{\omega_{(0)N} + |\omega_{(0)N}|}{2}f_{n+1} + \frac{\omega_{(0)N} - |\omega_{(0)N}|}{2}f_n,
\end{equation}
Here the one-sided differencing is employed and its direction is dictated by the signature of $\omega_{(0)n}$: the forward finite-differencing is adopted for $\omega_{(0)n}>0$, i.e., in the case for redshift whereas the backward differencing is employed for the blueshift case. 

The angular advection terms are finite-differenced just in a similar way, by switching the direction of one-sided differencing according to the signatures of $\omega_{(\ast)}$: employing $\mu_\nu=\mathrm{cos}\theta_\nu$, we write the $\theta_\nu$ advection term as
\begin{eqnarray}
&& -\frac{\partial}{\partial \mu_\nu}(\mathrm{sin}\theta_\nu f\omega_{(\theta_\nu)})\nonumber \\
\longrightarrow 
&&
-\frac{1}{(d\mu_\nu)_n}
\left[
\mathrm{sin}(\theta_\nu)_N \omega_{(\theta_\nu)N} f_N
- \mathrm{sin}(\theta_\nu)_{N-1} \omega_{(\theta_\nu)N-1} f_{N-1} \right];
\end{eqnarray}
with
\begin{equation}
\omega_{(\theta_\nu)N} f_N = \frac{\omega_{(\theta_\nu)N} + |\omega_{(\theta_\nu)N}|}{2}f_{n+1} + \frac{\omega_{(\theta_\nu)N} - |\omega_{(\theta_\nu)N}|}{2}f_n,
\end{equation}
The $\phi_\nu$ advection term is finite-differenced as 
\begin{eqnarray}
&& \frac{1}{\mathrm{sin}^2\theta_\nu}\frac{\partial}{\partial \phi_\nu}(f\omega_{(\phi_\nu)})\nonumber \\
\longrightarrow 
&&
\frac{1}{\mathrm{sin}^2(\theta_\nu)_{n_{\theta_\nu}}}\frac{1}{(d\phi_\nu)_{n_{\phi_\nu}}}
\left[
\omega_{(\phi_\nu){N_{\phi_\nu}}} f_{N_{\phi_\nu}}
- \omega_{(\phi_\nu){N_{\phi_\nu}}-1} f_{{N_{\phi_\nu}}-1} \right],
\end{eqnarray}
with 
\begin{equation}
\omega_{(\phi_\nu)N} f_N = \frac{\omega_{(\phi_\nu)N} - |\omega_{(\phi_\nu)N}|}{2}f_{n+1} + \frac{\omega_{(\phi_\nu)N} + |\omega_{(\phi_\nu)N}|}{2}f_n.
\end{equation}

\section{Geodesic Curves in the Schwarzschild Spacetime}
\label{sec:schwarzgeo}
We summarize how we calculate the geodesic curves in the Schwarzschild spacetime, employed as the reference in section \ref{sec:schwarzadv}. In the Schwarzschild (exterior) spacetime, the geodesic motion on the equatorial plane satisfies the following equation (see \S25.6 in \citet{MTW}, for example):
\begin{equation}
\label{eq:schwarzgeo}
\left(\frac{dr}{d\phi}\right)^2 
=
r^2\left(\frac{r^2}{b^2} + \frac{2GM}{c^2r} - 1\right),
\end{equation}
where $r$ and $\phi$ are the coordinate variables, and $b$ is the impact parameter. When the trajectory is on the meridional plane, one may replace the azimuth angle $\phi$ with the zenith angle $\theta$. The impact parameter $b$ can be expressed in terms of the radius $r_0$ and the zenith angle $(\theta_\nu)_0$ of a reference point on the geodesic curve as 
\begin{equation}
b=\frac{r_0\mathrm{sin}(\theta_\nu)_0}{1-2GM/(c^2r_0)},
\end{equation}
For numerical calculations, it is more useful to rewrite the above equation (\ref{eq:schwarzgeo}) as
\begin{equation}
\label{dudr}
\frac{d^2u}{d\phi^2}
=
\frac{3GMu^2}{c^2} - u,
\end{equation}
where the new variable is introduced as $u=1/r$. The second derivative of $u$ changes signature at the radius $r=3GM/c^2$, which corresponds to the radius of the photon sphere. In this work, equation (\ref{dudr}) is solved with the fourth-order explicit Runge-Kutta method by dividing it into the following two equations:
\begin{eqnarray}
&&
\frac{du}{d\phi} = v, \quad
\frac{dv}{d\phi} = \frac{3GMu^2}{c^2} - u.
\end{eqnarray}
We need for the angular advection tests in section \ref{sec:schwarzangadv} the zenith angle $\theta_\nu$ in momentum space along the geodesic, which is given in equation (13) in \citet{Shibata2014}:
\begin{equation}
\theta_\nu
= \mathrm{tan}^{-1}\left(\frac{p_{(2)}}{p_{(1)}}\right)
= \mathrm{tan}^{-1}\left(r\sqrt{1-\frac{2GM}{c^2r}}\frac{p^\theta}{p^r}\right),
\end{equation}
where $p_{(i)}$ is the momentum components for the tetrad basis and given as $p_{(i)} = p_{\mu}e^\mu_{(i)}$.

\section{Geodesic Curves in the Kerr Spacetime}
\label{sec:kerr}
For the Kerr spacetime, there is no simple differential equation to describe the geodesic curve unlike for the Schwarzschild spacetime. We hence solve the geodesic equation
\begin{equation}
\frac{d^2x^\mu}{d\lambda^2} = - \Gamma^\mu_{\alpha\beta}
\frac{dx^\alpha}{d\lambda}
\frac{dx^\beta}{d\lambda},
\end{equation}
where $\lambda$ is the affine parameter. Since we consider the geodesic curves only on the equatorial plane, we solve the equations only for $t$, $r$, and $\phi$. We employ the fourth order explicit Runge-Kutta for the following forms of equations:
\begin{eqnarray}
&&
\frac{dt}{d\lambda} = p^t,
\qquad
\frac{dp^t}{d\lambda} = - \Gamma^t_{\alpha\beta}p^\alpha p^\beta,
\nonumber \\
&&
\frac{dr}{d\lambda} = p^r,
\qquad
\frac{dp^r}{d\lambda} = - \Gamma^r_{\alpha\beta}p^\alpha p^\beta,
\nonumber \\
&&
\frac{d\phi}{d\lambda} = p^\phi,
\qquad
\frac{dp^\phi}{d\lambda} = - \Gamma^\phi_{\alpha\beta}p^\alpha p^\beta.
\end{eqnarray}
Here $p^\mu$ is the 4-momentum.

In the Kerr spacetime, the photon sphere for the prograde orbits with respect to the BH spin is given as 
\begin{equation}
r_\mathrm{prog} = \frac{2GM}{c^2}\left\{
1 + \mathrm{cos}\left[\frac{2}{3}\mathrm{cos}^{-1}\left(\frac{c^2|a|}{GM}\right)\right]
\right\},
\end{equation}
whereas for the retrograde orbits it is given as
\begin{equation}
r_\mathrm{ret} = \frac{2GM}{c^2}\left\{
1 + \mathrm{cos}\left[\frac{2}{3}\mathrm{cos}^{-1}\left(-\frac{c^2|a|}{GM}\right)\right]
\right\}.
\end{equation}
For the metric parameters, we employ the same values as the advection tests in section \ref{sec:kerrtest}; with $M=5M_\odot$ and $a=0.5GM/c^2$.

\begin{figure}[t]
  \includegraphics[width=18cm]{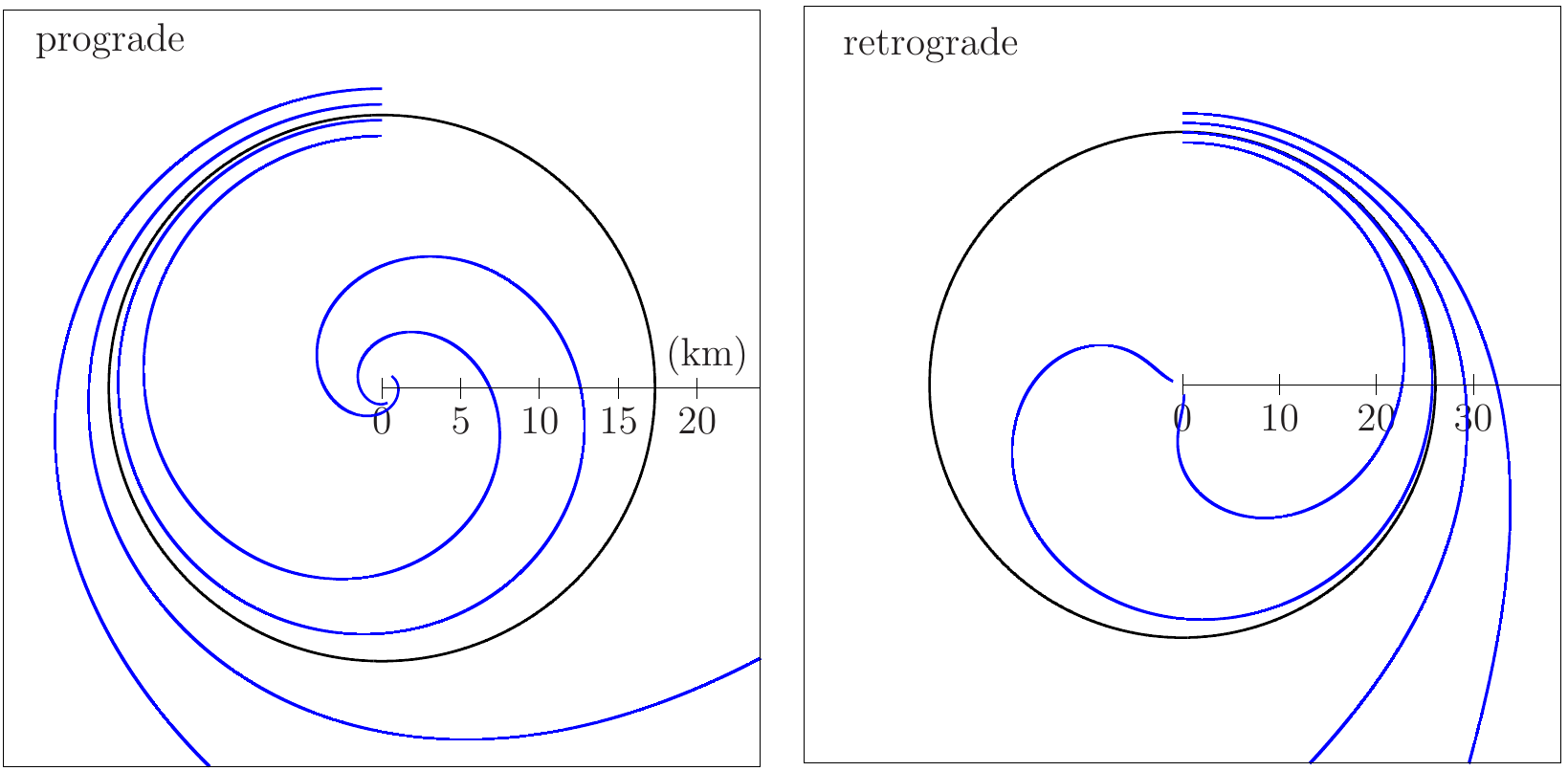}
  \caption{Some geodesic curves on the equatorial plane in the Kerr spacetime. The mass and spin of BH are set to $M=5M_\odot$ and $a=0.5GM/c^2$, respectively. Left: prograde geodesic curves with $p^r = 0$ at $r = 16$, $17$, $18$, $19\,{\rm km}$. Right: retrograde geodesic curves with $p^r = 0$ at $r = 25$, $26$, $27$, $28\,{\rm km}$. The photon spheres are indicated as the black circles, the radius of which is $17.33\,{\rm km}$ and $26.07\,{\rm km}$ for the prograde and retrograde orbits, respectively.}
  \label{Fig:kerrgeo}
\end{figure}

Figure \ref{Fig:kerrgeo} shows some geodesic curves on the equatorial plane. The left panel exhibits four prograde geodesic curves with $p^r=0$ at $r=16$, $17$, $18$ and $19\,{\rm km}$. The first two radii are smaller than that of the photon sphere, which is $r=17.33\,{\rm km}$ in the present case whereas the last two are larger. The photon sphere is indicated with the black circle in the figure. The right panel, on the other hand, presents some retrograde geodesic curves with $p^r = 0$ at $r=25$, $26$, $27$, and $28\,{\rm km}$. The photon sphere has the radius of $r=26.07\,{\rm km}$ in this case. It is found in both panels that the geodesic curves are confined either inside or outside the photon sphere. 

For the advection tests in the Kerr spacetime in section \ref{sec:kerrtest}, we need the zenith angle $\theta_{\nu}$ of the 4-momentum on the geodesic curve. Just as for the Schwarzschild spacetime, it is obtained from
\begin{equation}
\theta_\nu
= \mathrm{tan}^{-1}\left(\frac{p_{(3)}}{p_{(1)}}\right) = \mathrm{tan}^{-1}\left(\frac{p_re^r_{(3)}+p_\phi e^\phi_{(3)}}{p_re^r_{(1)}}\right).
\end{equation}


\bibliography{sample63}{}
\bibliographystyle{aasjournal}



\end{document}